\documentclass[journal,twocolumn,10pt]{IEEEtran}
\usepackage{amsmath,amssymb}
\usepackage{paralist}
\usepackage{bm}
\usepackage{epsfig, epstopdf}
\usepackage{graphics}
\usepackage{multirow, rotating}
\usepackage{pstricks}
\usepackage{subfigure}
\usepackage{booktabs}

\newtheorem{theorem}{Theorem}

\newtheorem{definition}{Definition}

\renewcommand{\S}{\mathcal{S}}

\newcommand{\Neighb}{\mathcal{N}}
\newcommand{\C}{\mathbb{C}}
\newcommand{\R}{\mathbb{R}}

\newcommand{\coord}[1]{\mathbf{#1}}
\newcommand{\argmin}{\mathbf{arg min}}

\newcommand{\DSPG}{$\mbox{DSP}_{\mbox{\scriptsize G}}$}

\newcommand{\mypar}[1]{{\bf #1.}}

\newcommand{\Nodes}{\mathcal{V}}
\newcommand{\Adj}{\mathbf{A}}
\newcommand{\Adjn}{\mathbf{A}^{\scriptsize\textrm{norm}}}

\DeclareMathOperator{\TV}{TV}
\DeclareMathOperator{\TVG}{TV_G}
\DeclareMathOperator{\TVL}{TV_L}
\DeclareMathOperator{\ind}{i}
\DeclareMathOperator{\Vm}{\mathbf{V}}
\DeclareMathOperator{\Jm}{\mathbf{J}}
\DeclareMathOperator{\Eig}{\mathbf{\Lambda}}

\DeclareMathOperator{\FT}{\mathbf{F}}

\DeclareMathOperator{\Sf}{S}

\DeclareMathOperator{\Circ}{\mathbf{C}}
\DeclareMathOperator{\DFT}{\mathbf{DFT}}
\DeclareMathOperator{\Id}{\mathbf{I}}

\DeclareMathOperator{\Cm}{\mathbf{C}}
\DeclareMathOperator{\Dm}{\mathbf{D}}

\DeclareMathOperator{\Lm}{\mathbf{L}}

\newlength{\whitespacelength}
\title{Discrete Signal Processing on Graphs:\\ Frequency Analysis}
\author{Aliaksei Sandryhaila$^1$ and Jos\'{e} M.~F. Moura$^1$
}

\author{Aliaksei Sandryhaila,~\IEEEmembership{Member,~IEEE}
    and Jos\'{e} M.~F. Moura,~\IEEEmembership{Fellow,~IEEE}
  \thanks{This work was supported in part by AFOSR grant FA95501210087. A.~Sandryhaila and J.~M.~F.~Moura are with the Department of Electrical and Computer
    Engineering, Carnegie Mellon University, Pittsburgh, PA 15213-3890. Ph:~(412)268-6341; fax:~(412)268-3890. Email: asandryh@andrew.cmu.edu, moura@ece.cmu.edu.}
}

\begin{document}

\maketitle

\begin{abstract}
Signals and datasets that arise in physical and engineering applications, as well as social, genetics, biomolecular, and many other domains, are becoming increasingly larger and more complex.
In contrast to traditional time and image signals, data in these domains are supported by arbitrary graphs. Signal processing on graphs extends concepts and techniques from traditional signal processing to data indexed by generic graphs. This paper studies the concepts of low and high frequencies on graphs, and low-, high-, and band-pass graph filters. In traditional signal processing, there concepts are easily defined because of a natural frequency ordering that has a physical interpretation. For signals residing on graphs, in general, there is no obvious frequency ordering.
We propose a definition of total variation for graph signals that naturally leads to a frequency ordering on graphs and defines low-, high-, and band-pass graph signals and filters.
We study the design of graph filters with specified frequency response, and illustrate our approach with applications to sensor malfunction detection and data classification.
\end{abstract}

\textbf{Keywords}:
Signal processing on graphs, graph filter, total variation, filter design, regularization, low pass, high pass, band pass.


\section{Introduction}
\label{sec:Intro}

Signals indexed by graphs arise in many applications, including the analysis of preferences and opinions in social and economic networks~\cite{Jackson:08,Newman-2010,Easley:10}; research in collaborative activities, such as paper co-authorship and citations~\cite{Newman:04}; topics and relevance of documents in the World Wide Web~\cite{Adamic:05,Brin:98}; customer preferences for service providers; measurements from sensor networks; interactions in molecular and gene regulatory networks; and many others.

\emph{Signal processing on graphs} extends the classical discrete signal processing (DSP) theory for time signals and images~\cite{Oppenheim:99} to signals indexed by vertices of a graph. There are two basic approaches to signal processing on graphs. The first one uses the graph Laplacian matrix as its basic building block (see a recent review~\cite{Shuman:13} and references therein). The second approach adopts the adjacency matrix of the underlying graph as its fundamental building block~\cite{Sandryhaila:13,Sandryhaila:13a,Sandryhaila:13b}.
Both frameworks define several signal processing concepts similarly, but the difference in their foundation leads to different techniques for signal analysis and processing.

Methods for Laplacian-based graph signal analysis emerged from research on the spectral graph theory~\cite{Chung:96} and manifold discovery and embedding~\cite{Tenenbaum:00,Roweis:00}.
Implicitly or explicitly, in these works graphs discretize continuous high-dimensional manifolds from $\R^M$: graph vertices sample a manifold and connect to nearest neighbors as determined by their geodesic distances with respect to the underlying manifold. In this setting, the graph Laplacian operator is the discrete counterpart to the continuous Laplace-Beltrami operator on a manifold~\cite{Chung:96,Coifman:05a}.

This connection is propagated conceptually to Laplacian-based methods for signal processing on graphs. For example, the graph Fourier transform defined and considered in~\cite{Shuman:13}, as well as~\cite{Hammond:11,Narang:12,Agaskar:13,Ekambaram:13,Ekambaram:13a,Zhu:12}, expands graph signals in the eigenbasis of the graph Laplacian. This parallels the classical Fourier transform that expands signals into the basis of complex exponentials that are eigenfunctions of the one-dimensional Laplace operator -- the negative second order derivative operator~\cite{Shuman:13}. The frequencies are the eigenvalues of the Laplace operator. Since the operator is symmetric and positive semi-definite, graph frequencies are real-valued and hence totally ordered. So, just like for time signals, the notions of low and high frequencies are easily defined in this model. However, due to the symmetry and positive semi-definiteness of the operator, the Laplacian-based methods are only applicable to undirected graphs with real, non-negative weights.

In~\cite{Sandryhaila:13,Sandryhaila:13a,Sandryhaila:13b} we take a different route.
 Our approach is motivated by the algebraic signal processing~(ASP) theory introduced in~\cite{Pueschel:03a,Pueschel:05e,Pueschel:08a,Pueschel:08b,Pueschel:08c}; see also~\cite{Pueschel:05c,Pueschel:05d,Pueschel:07,Sandryhaila:12,Sandryhaila:11a} for additional developments. In ASP, the shift is the elementary non-trivial filter that generates, under an appropriate notion of shift invariance, all linear shift-invariant filters for a given class of signals. The key insight in~\cite{Sandryhaila:13} to build the theory of signal processing on graphs is to identify the shift operator. We adopted the weighted adjacency matrix of the graph as the shift operator and then developed appropriate concepts of $z$-transform, impulse and frequency response, filtering, convolution, and Fourier transform. In particular, the graph Fourier transform in this framework expands a graph signal into a basis of eigenvectors of the adjacency matrix,
and the corresponding spectrum is given by the eigenvalues of the adjacency matrix. This contrasts with the Laplacian-based approach, where Fourier transform and spectrum are defined by the eigenvectors and eigenvalues of the graph Laplacian, while in our approach the eigenvectors and eigenvalues to expand graph signals are obtained from the spectrum of the adjacency matrix.

The association of the graph shift with the adjacency matrix is natural and has multiple intuitive interpretations. The graph shift is an elementary filter,
and its output is a graph signal with the value at vertex~$n$ given approximately by a weighted linear combination of the input signal values at neighbors of~$n$~\cite{Sandryhaila:13}. With appropriate edge weights, the graph shift can be interpreted as a (minimum mean square) first-order linear predictor~\cite{Pueschel:05e,Sandryhaila:13}.
Another interpretation of the graph filter comes from Markov chain theory~\cite{Papoulis:02}, where the adjacency matrix represents the one-step transition probability matrix of the chain governing its dynamics. Finally, the graph shift can also be seen as a stencil approximation of the first-order derivative on the graph\footnote{This analogy is more intuitive to understand if the graph is regular.}.

The last interpretation of the graph shift contrasts with the corresponding interpretation of the graph Laplacian: the adjacency matrix is associated with a first-order differential operator, while the Laplacian, if viewed as a shift, is associated with a second-order differential operator.  In the one-dimensional case, the eigenfunctions for both, the first order and second order differential operators, are complex exponentials, since
\begin{align}
\label{eqn:1Dtimefrequency}
\frac{1}{2\pi j}\frac{d}{dt} e^{2\pi j ft}=f e^{2\pi j ft}.
\end{align}
Interpreting the Laplacian as a shift introduces an even symmetry assumption into the corresponding signal model, and for one-dimensional signals~\cite{Pueschel:08b}, this model assumes that the signals are defined on lines of an image (undirected line graphs) rather than on a time line (directed line graphs). The use of the adjacency matrix as the graph shift does not impose such assumptions, and the corresponding framework can be used for arbitrary signals indexed by general graphs, regardless whether these graphs have undirected or directed edges with real or complex, non-negative or negative weights.

This paper is concerned with defining low and high frequencies and low-, high-, and band-pass graph signals and filters on generic graphs. In traditional discrete signal processing~(DSP), these concepts have an intuitive interpretation, since the frequency contents of time series and digital images are described by complex or real sinusoids that oscillate at different rates~\cite{Mallat:08}. The oscillation rates provide a physical notion of ``low'' and ``high'' frequencies: low-frequency components oscillate less and high-frequency ones oscillate more. However, these concepts do not have a similar interpretation on graphs, and it is not obvious how to order graph frequencies to describe the low- and high-frequency contents of a graph signal.

We present an ordering of the graph frequencies that is based on how ``oscillatory'' the spectral components are with respect to the indexing graph, i.e., how much they change from a node to  neighboring nodes. To quantify this amount, we introduce the \emph{graph total variation} function that measures how much signal samples (values of a graph signal at a node) vary in comparison to neighboring samples. This approach is analogous to the classical DSP theory, where the oscillations in time and image signals are also quantified by appropriately defined total  variations~\cite{Mallat:08}. In Laplacian-based graph signal processing~\cite{Shuman:13}, the authors choose to order frequencies based on a quadratic form rather than on the total variation of the graph signal.  Once we have an ordering of the frequencies based on the graph total variation function, we define the notions of low and high frequencies, low-, high-, and band-pass graph signals, and low-, high-, and band-pass graph filters.
We demonstrate that these concepts can be used effectively in sensor networks analysis and classification of hyperlinked political blogs. In our experiments, we show that naturally occurring graph signals, such as measurements of physical quantities collected by sensor networks or labels for political blogs in a  dataset, tend to be low-frequency graph signals, while anomalies in sensor measurements or missing data labels can amplify high-frequency parts of the signals. We demonstrate how these anomalies can be detected using appropriately designed high-pass graph filters, and how unknown parts of graph signals can be recovered with appropriately designed regularization techniques.

\textbf{Summary of the paper.} In Section~\ref{sec:DSPG}, we present the notation and review from~\cite{Sandryhaila:13} the basics of discrete signal processing on graphs~(\DSPG). In Section~\ref{sec:TV}, we define the local and total variation for graph signals. In Section~\ref{sec:LF_HF}, we use the proposed total variation to impose an ordering on frequency components from lowest to highest.
In Section~\ref{sec:FilterDesign}, we discuss low-, high-, and band-pass graph filters and their design. In Section~\ref{sec:Applications}, we illustrate these concepts with applications to corrupted measurement detection in sensor networks and data classification, and provide experimental results for real-world datasets. Finally, Section~\ref{sec:Conclusions} concludes the paper.

\section{Discrete Signal Processing on Graphs}
\label{sec:DSPG}

In this section, we briefly review notation and main concepts of the \DSPG\ framework that are relevant to our work in this paper. A complete introduction to the theory can be found in~\cite{Sandryhaila:13,Sandryhaila:13a,Sandryhaila:13b}.

\subsection{Graph Signals}

Signal processing on graphs is concerned with the analysis and processing of datasets in which data elements can be connected to each other according to some relational property. This relation is expressed though a graph $G=(\Nodes,\Adj)$, where $\Nodes=\{v_0,\ldots,v_{N-1}\}$ is a set of nodes and $\Adj$ is a weighted adjacency matrix of the graph. Each data element corresponds to node $v_n$ (we also say the data element is \emph{indexed} by $v_n$), and each weight $\Adj_{n,m}\in\C$ of a directed edge from $v_m$ to $v_n$ reflects the degree of relation of the $m$th data element to the $n$th one. Node~$v_m$ is an in-neighbor of~$v_n$ and $v_n$ is an out-neighbor of~$v_m$ if $\Adj_{n,m}\neq 0$. All in-neighbors of $v_n$ form its \emph{in-neighborhood}, and we denote a set of their indices as
$\Neighb_n = \{ m \mid \Adj_{n,m}\neq 0\}$. If the graph is undirected, the relation goes both ways, $\Adj_{n,m}=\Adj_{m,n}$, and the nodes are neighbors.

Using this graph, we refer to the dataset as a \emph{graph signal}, which is defined as a map
\begin{eqnarray}
\nonumber
\coord{s} & : & \Nodes\,\rightarrow\,\C, \\
\label{eq:s}
&& v_n\,\mapsto\,s_n.
\end{eqnarray}
We assume that each dataset element $s_n$ is a complex number. Since each signal is isomorphic to a complex-valued vector with $N$ elements, we write graph signals as vectors
$$
\coord{s} = \begin{bmatrix} s_0 & s_1 & \ldots & s_{N-1} \end{bmatrix}^T \in \C^N.
$$
However, we emphasize that each element $s_n$ is indexed by node $v_n$ of a given representation graph $G=(\Nodes,\Adj)$, as defined by~\eqref{eq:s}. The space $\S$ of graph signals~\eqref{eq:s} is isomorphic to $\C^N$, and its dimension is $\dim\S=N$.

\subsection{Graph Filters}
\label{sec:GraphFilters}

In general, a \emph{graph filter} is a system $\coord{H}(\cdot)$ that takes a graph signal $\coord{s}$ as an input, processes it, and produces another graph signal $\coord{\tilde{s}}=\coord{H}(\coord{s})$ as an output.
A basic non-trivial filter defined on a graph $G=(\Nodes, \Adj)$, called the \emph{graph shift}, is a local operation that replaces a signal value $s_n$ at node $v_n$ with the linear combination of values at the neighbors of node $v_n$
\begin{equation}
\label{eq:graph_shift_n}
\tilde{s}_n = \sum_{m\in\Neighb_n}{\Adj_{n,m} s_m}.
\end{equation}
Hence, the output of the graph shift is given by the product of the input signal with the adjacency matrix of the graph:
\begin{equation}
\label{eq:graph_shift}
\coord{\tilde{s}}=
\begin{bmatrix} \tilde{s}_0 & \ldots & \tilde{s}_{N-1} \end{bmatrix}^T
=\Adj\coord{s}.
\end{equation}
The graph shift is the basic building block in \DSPG.

All linear, shift-invariant\footnote{
Filters are \emph{linear} if for a linear combination of inputs they produce the same linear combination of outputs. Filters are \emph{shift-invariant} if the result of consecutive processing of a signal by multiple graph filters does not depend on the order of processing; i.e., shift-invariant filters commute with each other.}
graph filters in \DSPG\ are polynomials in the adjacency matrix $\Adj$ of the form~\cite{Sandryhaila:13}
\begin{equation}
\label{eq:h_A}
h(\Adj) = h_0\Id + h_1\Adj + \ldots + h_{L}\Adj^{L}.
\end{equation}
The output of the filter~\eqref{eq:h_A} is the signal
$$
\coord{\widetilde{s}}=\coord{H}(\coord{s}) = h(\Adj)\coord{s}.
$$

Linear, shift-invariant graph filters possess a number of useful properties. They have at most $L\leq N_\Adj$ taps $h_\ell$, where $N_\Adj=\deg m_\Adj(x)$ is the degree of the minimal polynomial\footnote{The minimal polynomial of $\Adj$ is the unique monic polynomial of the smallest degree that annihilates $\Adj$, i.e., $m_\Adj(\Adj) = 0$~\cite{Lancaster:85,Gantmacher:59}.} $m_\Adj(x)$ of $\Adj$. If a graph filter~\eqref{eq:h_A} is invertible, i.e., matrix $h(\Adj)$ is non-singular, then its inverse is also a graph filter $g(\Adj)=h(\Adj)^{-1}$ on the same graph $G=(\Nodes,\Adj)$. Finally, the space of graph filters is an \emph{algebra}, i.e., a vector space that is simultaneously a ring.

These properties guarantee that multiplying $\Adj$ by any non-zero constant does not change the set of corresponding linear, shift-invariant graph filters.
In particular, we can define the normalized matrix
\begin{equation}
\label{eq:Adj_normalized}
\Adjn = \frac{1}{|\lambda_{\scriptsize \textrm{max}}|} \Adj,
\end{equation}
where $\lambda_{\scriptsize \textrm{max}}$ denotes the eigenvalue of $\Adj$ with the largest magnitude, i.e.,
\begin{equation}
\label{eq:lambda_max}
|\lambda_{\scriptsize \textrm{max}}|\geq |\lambda_m|
\end{equation}
for all $0\leq m \leq M-1$. The normalized matrix~\eqref{eq:Adj_normalized} has spectral norm $|| \Adjn ||_2 = 1$ that guarantees
that the shifted signal is not scaled up, since $\left|\left| \Adjn \coord{s} \right|\right| / \left|\left| \coord{s} \right|\right| \leq 1$,
and ensures the numerical stability of computing with graph filters $h(\Adjn)$.
In this paper, we use the graph shift $\Adjn$ instead of $\Adj$ when these guarantees are required.

\subsection{Graph Fourier Transform}

In general, a Fourier transform performs the expansion of a signal into a \emph{Fourier basis} of signals that are invariant to filtering. In the \DSPG\ framework, a graph Fourier basis corresponds to the Jordan basis of the graph adjacency matrix $\Adj$ (the Jordan decomposition is reviewed in Appendix~A). Following the DSP notation, distinct eigenvalues $\lambda_0,\lambda_1,\ldots, \lambda_{M-1}$ of the adjacency matrix $\Adj$ are called the \emph{graph frequencies} and form the \emph{spectrum} of the graph, and the Jordan eigenvectors that correspond to a frequency $\lambda_m$ are called the \emph{frequency components} corresponding to the $m$th frequency. Since multiple eigenvectors can correspond to the same eigenvalue, in general, a graph frequency can have multiple graph frequency components associated with it.

As reviewed in Appendix~A, Jordan eigenvectors form the columns of the matrix $\Vm$ in the Jordan decomposition~\eqref{eq:Jordan_decomposition}
$$
\Adj=\Vm\Jm\Vm^{-1}.
$$
Hence, the \emph{graph Fourier transform} of a graph signal $\coord{s}$ is
\begin{equation}
\label{eq:graph_FT}
\coord{\widehat{s}} = \FT \coord{s},
\end{equation}
where $\FT=\Vm^{-1}$ is the graph Fourier transform matrix. The values $\widehat{s}_n$ of the signal's graph Fourier transform~\eqref{eq:graph_FT} characterize the \emph{frequency content} of the signal $\coord{s}$.

The \emph{inverse graph Fourier transform} is given by
\begin{equation}
\label{eq:graph_FT_inverse}
\coord{s}= \FT^{-1}\coord{\widehat{s}} = \Vm\coord{\widehat{s}}.
\end{equation}
It reconstructs the original signal from its frequency contents by constructing a linear combination of frequency components weighted by the signal's Fourier transform coefficients.

\subsection{Frequency Response}

The graph Fourier transform~\eqref{eq:graph_FT} also allows us to characterize the effect of a filter on the frequency content of an input signal. As follows from~\eqref{eq:h_A} and~\eqref{eq:graph_FT}, as well as~\eqref{eq:Jordan_decomposition} in Appendix~A,
\begin{equation}
\label{eq:conv_theorem}
\widetilde{\coord{s}} = h(\Adj)\coord{s} = \FT^{-1} h(\Jm) \FT \coord{s}
\,\,\,\Leftrightarrow\,\,\,
\FT\widetilde{\coord{s}} = h(\Jm) \coord{\widehat{s}}.
\end{equation}
Hence, the frequency content of the output signal is obtained by multiplying the frequency content of the input signal by the block diagonal matrix
\begin{equation*}
h(\Jm) = \begin{bmatrix}
h(\Jm_{r_{0,0}}(\lambda_0)) \\
& \ddots \\
&& h(\Jm_{r_{M-1,D_M-1}}(\lambda_{M-1}))
\end{bmatrix}.
\end{equation*}
We call this matrix the \emph{graph frequency response} of the filter $h(\Adj)$, and denote it as
\begin{equation}
\label{eq:frequency_response}
\widehat{h(\Adj)} = h(\Jm).
\end{equation}

Notice that~\eqref{eq:conv_theorem} extends the \emph{convolution theorem} from classical signal processing~\cite{Oppenheim:99} to graphs, since filtering a signal on a graph is equivalent in the frequency domain to multiplying the signal's spectrum by the frequency response of the filter.

\subsection{Consistency with the Classical DSP}

\begin{figure}
  \begin{center}
    \includegraphics[scale=0.6]{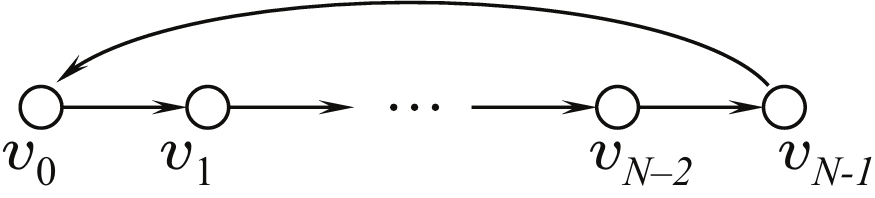}
  \end{center}
  \vspace{-2mm}
\caption{\label{fig:graph_time} Traditional graph representation for a finite discrete periodic time series of length $N$.}
\vspace{-5mm}
\end{figure}

The \DSPG\ framework is consistent with the classical DSP theory. Finite (or periodic) time series can be represented by the directed cycle graph shown in Fig.~\ref{fig:graph_time}, see~\cite{Pueschel:08a,Sandryhaila:13}. The direction of the edges represents the flow of time from past to future, and the edge from the last vertex $v_{N-1}$ to~$v_0$ captures the periodic signal extension $s_N = s_0$ for time series. The adjacency matrix of the graph in Fig.~\ref{fig:graph_time} is the $N\times N$ cyclic permutation matrix
\begin{equation}
\label{eq:circulant}
\Adj = \Circ = \begin{bmatrix}
&&&1\\
1\\
& \ddots \\
&& 1
\end{bmatrix}.
\end{equation}
Substituting~\eqref{eq:circulant} into the graph shift~\eqref{eq:graph_shift} yields the standard time delay
\begin{equation}
\label{eq:time_delay}
\widetilde{s}_n = s_{n-1 \mod N}.
\end{equation}

The matrix~\eqref{eq:circulant} is diagonalizable. Its eigendecomposition, which coincides with the Jordan decomposition~\eqref{eq:Jordan_decomposition}, is
\begin{equation*}
\Circ =
\frac{1}{N}
\DFT_N^{-1}
\begin{bmatrix}
e^{-j\frac{2\pi \cdot 0}{N}} \\
& \ddots \\
&& e^{-j\frac{2\pi\cdot (N-1)}{N}}
\end{bmatrix}
\DFT_N,
\end{equation*}
where $\DFT_N$ is the discrete Fourier transform matrix. Thus, as expected, the graph Fourier transform for signals indexed by the graph in Fig.~\ref{fig:graph_time} is $\FT=\DFT_N$, and the corresponding frequencies are\footnote{In DSP, the ratio $\frac{2\pi}{N}n$ in the exponent~\eqref{eq:exponent} sometimes is also called frequency. In this case, the frequencies are $N$ real numbers between $0$ and $2\pi$. However, to remain consistent with the discussion in this paper, we refer to the exponentials~\eqref{eq:exponent} as frequencies, and view them as complex numbers of magnitude $1$ residing on the unit circle in the complex plane.}, for $0\leq n < N$,
\begin{equation}
\label{eq:exponent}
e^{-j\frac{2\pi}{N}n}.
\end{equation}

\section{Total Variation on Graphs}
\label{sec:TV}

In this section, we define he total variation on graph signals that is based on the concept of graph shift.

In classical DSP, the \emph{total variation} (TV) of a discrete signal is defined as the sum of magnitudes of differences between two consecutive signal samples~\cite{Mallat:08}:
\begin{equation}
\label{eq:time_TV_infinite}
\TV(\coord{s}) = \sum_n \big|s_n-s_{n-1}\big|.
\end{equation}
For a finite time series, the periodicity condition $s_{n} = s_{n\mod N}$ yields a modified definition
\begin{equation}
\label{eq:time_TV}
\TV(\coord{s}) = \sum_{n=0}^{N-1} \big|s_n - s_{n-1 \mod N}\big|.
\end{equation}

The total variation~\eqref{eq:time_TV_infinite} and~\eqref{eq:time_TV} for time series or space signals, such as images, has an intuitive interpretation: it compares how the signal varies with time or space. These concepts lie at the heart of many applications of DSP, including signal regularization and denoising~\cite{Mallat:08,VetterliKovacevic:95}, image compression~\cite{Bovik:05} and others.

The variation~\eqref{eq:time_TV} compares two consecutive signal samples and calculates a cumulative magnitude of the signal change over time. In terms of the time shift~\eqref{eq:time_delay}, we can say that the total variation compares a signal $\coord{s}$ to its shifted version: the smaller the difference between the original signal and the shifted one, the lower the signal's variation. Using the cyclic permutation matrix~\eqref{eq:circulant}, we can write~\eqref{eq:time_TV} as
\begin{equation}
\label{eq:time_TV_shift}
\TV(\coord{s}) = \left|\left| \coord{s} - \Circ \coord{s}\right|\right|_1.
\end{equation}
The total variation~\eqref{eq:time_TV_shift} measures the difference between the signal samples at each vertex and at its neighbor on the graph that represents finite time series in Fig.~\ref{fig:graph_time}.

The \DSPG\ generalizes the DSP theory from lines and regular lattices to arbitrary graphs. Hence, we extend~\eqref{eq:time_TV_shift} to an arbitrary graph $G=(\Nodes,\Adj)$ by defining the total variation on a graph as a measure of similarity between a graph signal and its shifted version~\eqref{eq:graph_shift}:

\begin{definition}[Total Variation on Graphs]
\label{def:graphtotalvariation}
The \emph{total variation on a graph} ($\TVG$) of a graph signal $\coord{s}$ is defined as
\begin{equation}
\label{eq:TV_graph}
\TVG(\coord{s}) = \left|\left|\coord{s} - \Adjn \coord{s}\right|\right|_1.
\end{equation}
\end{definition}
The definition uses the normalized adjacency matrix $\Adjn$ to guarantee that the shifted signal is properly scaled for comparison with the original signal, as discussed in Section~\ref{sec:GraphFilters}.

The intuition behind Definition~\ref{def:graphtotalvariation} is supported by the underlying mathematical model.
Similarly to the calculus on discrete signals that defines the discretized derivative as $\nabla_n(\coord{s}) = s_n - s_{n-1}$~\cite{Mallat:08},
in \DSPG\ the derivative (and the gradient) of a graph signal at the $n$th vertex is defined by the graph shift $\Adjn$ as
\begin{equation}
\label{eq:gradient}
\frac{d \coord{s}}{d v_n} = \nabla_n(\coord{s}) = s_n - \sum_{m\in\Neighb_n}{\Adjn_{n,m} s_m} .
\end{equation}
The \emph{local variation} of the signal at vertex $v_n$ is the magnitude $\left| \nabla_n(\coord{s}) \right|$ of the corresponding gradient,
and the \emph{total variation} is the sum of local variations for all vertices~\cite{Mallat:08,Shuman:13}.
In particular, if we define the discrete $p$-Dirichlet form
\begin{equation}
\label{eq:Dirichlet}
\Sf_p(\coord{s}) = \frac{1}{p} \sum_{n=0}^{N-1} \left| \nabla_n(\coord{s}) \right|^p,
\end{equation}
then for $p=1$ the form
\begin{eqnarray}
\label{eq:Dirichlet_1}
\Sf_1(\coord{s}) &=& \sum_{n=0}^{N-1} \left| \nabla_n(\coord{s}) \right| \\
\nonumber
&=& \sum_{n=0}^{N-1} \left| s_n - \sum_{m\in\Neighb_n}{\Adjn_{n,m} s_m}  \right| \\
\nonumber
&=& \left|\left| \coord{s} - \Adjn\coord{s} \right|\right|_1
\end{eqnarray}
defines the total variation of the graph signal $\coord{s}$. It coincides with Definition~\ref{def:graphtotalvariation}.

The total variation defined through the $1$-Dirichlet form~\eqref{eq:Dirichlet_1} depends on the definition of the signal gradient at a graph vertex. For finite time DSP, the gradient is defined by the discretized derivative $\nabla_n(\coord{s}) = s_n - s_{n-1}$~\cite{Mallat:08} and yields the total variation~\eqref{eq:time_TV_shift}. The \DSPG\ extends the notion of the shift to~\eqref{eq:graph_shift_n}, which leads to the gradient~\eqref{eq:gradient} and the total variation~\eqref{eq:TV_graph}.

\textbf{Remark.} In~\cite{Shuman:13}, the frequencies are ordered using a 2-Dirichlet form, i.e., a quadratic function.
%
%
%

\section{Low and High Frequencies on Graphs}
\label{sec:LF_HF}

In this section, we use the total variation~\eqref{eq:TV_graph} to introduce an ordering on frequencies that leads to the definition of low and high frequencies on graphs.
We demonstrate that this ordering is unique for graphs with real spectra and not unique for graphs with complex spectra.

\subsection{Variation of the Graph Fourier Basis}

As discussed in Section~\ref{sec:DSPG}, the graph Fourier basis for an arbitrary graph is given by the Jordan basis of the adjacency matrix $\Adj$. Consider an eigenvalue $\lambda$ of $\Adj$, and let $\coord{v}=\coord{v}_0,\coord{v}_1,\ldots,\coord{v}_{R-1}$ be a Jordan chain of generalized eigenvectors that corresponds to this eigenvalue. Let the indicator function
$$
\ind_r = \begin{cases}
0,&\,\,\, r=0 \\
1,&\,\,\, 1\leq r < R
\end{cases}
$$
specify whether $\coord{v}_r$ is a proper eigenvector of $\Adj$ or a generalized one. Then we can write the condition~\eqref{eq:generalized_eigenvector_condition} on generalized eigenvectors (see Appendix~A) as
\begin{equation}
\label{eq:generalized_eigenvector_condition_indicator}
\Adj\coord{v}_r = \lambda \coord{v}_r + \ind_r \coord{v}_{r-1}.
\end{equation}

Using~\eqref{eq:generalized_eigenvector_condition_indicator}, we write the total variation~\eqref{eq:TV_graph} of the generalized eigenvector $\coord{v}_r$ as
\begin{eqnarray}
\label{eq:TV_generalized}
\TVG(\coord{v}_r)
&=& \left|\left|\coord{v}_r - \Adjn\coord{v}_r\right|\right|_1 \\
\nonumber
&=&  \left|\left|\coord{v}_r - \frac{1}{|\lambda_{\scriptsize \textrm{max}}|}\Adj\coord{v}_r\right|\right|_1 \\
\nonumber
&=& \left|\left|\coord{v}_r - \frac{\lambda}{|\lambda_{\scriptsize \textrm{max}}|}\coord{v}_r - \frac{\ind_r}{|\lambda_{\scriptsize \textrm{max}}|}\coord{v}_{r-1}\right|\right|_1 .
\end{eqnarray}
In particular, when $\coord{v}_r$ is a proper eigenvector of $\Adj$, i.e. $r=0$ and $\coord{v}_0=\coord{v}$, we have $\ind_0=0$. In this case, it follows from~\eqref{eq:TV_generalized} that the total variation of the eigenvector $\coord{v}$ is
\begin{equation}
\label{eq:TV_eigenvector}
\TVG(\coord{v}) = \left|1-\frac{\lambda}{|\lambda_{\scriptsize \textrm{max}}|}\right|  \left| \left|\coord{v} \right|\right|_1 .
\end{equation}

When a frequency component is a proper eigenvector of the adjacency matrix $\Adj$, its total variation~\eqref{eq:TV_eigenvector} is determined by the corresponding eigenvalue, since we can scale
all eigenvectors to have the same $\ell_1$-norm. Moreover, all proper eigenvectors corresponding to the same eigenvalue have the same total variation. However, when a frequency component is not a proper eigenvector, we must use~\eqref{eq:TV_generalized} to compare its variation with other frequency components.
Finally, it follows from~\eqref{eq:lambda_max} for $\left| \left|\coord{v} \right|\right|_1=1$ that
\begin{equation}
\label{eq:TV_bound}
\TVG(\coord{v}) = \left|1-\frac{\lambda}{|\lambda_{\scriptsize \textrm{max}}|}\right| \leq 1 + \left|\frac{\lambda}{|\lambda_{\scriptsize \textrm{max}}|}\right| \leq 2.
\end{equation}
Hence, the total variation of a normalized proper eigenvector is a real number between $0$ and $2$.

\subsection{Frequency Ordering}
\label{sec:Frequency}
%
%
The total variation of the Fourier basis, given by~\eqref{eq:TV_generalized} and~\eqref{eq:TV_eigenvector}, allows us to order the graph frequency components in the order of increasing variation. Following DSP convention, we call frequency components with smaller variations \emph{low} frequencies and components with higher variations \emph{high} frequencies.

Here, we determine the frequency ordering induced by the total variation~\eqref{eq:TV_eigenvector} for graphs that have diagonalizable adjacency matrices, i.e., only have proper eigenvectors. This ordering can be similarly extended to graphs with non-diagonalizable adjacency matrices using the generalized eigenvector variation~\eqref{eq:TV_generalized}.

The following theorem establishes the relative ordering of two distinct real frequencies.

\begin{theorem}
\label{thm:ordering_real}
Consider two distinct real eigenvalues $\lambda_m,\lambda_n\in\R$ of the adjacency matrix $\Adj$ with corresponding eigenvectors $\coord{v}_m$ and $\coord{v}_n$.
If the eigenvalues are ordered as
\begin{equation}
\label{eq:real_lambda_condition}
\lambda_m < \lambda_n,
\end{equation}
then the total variations of their eigenvectors satisfy
\begin{equation}
\label{eq:frequency_ordering_real}
\TVG(\coord{v}_m) > \TVG(\coord{v}_n).
\end{equation}
\end{theorem}
\begin{IEEEproof}
Since the eigenvalues are real, it follows from~\eqref{eq:real_lambda_condition} that the difference between the total variations of the two eigenvectors satisfies
\begin{eqnarray*}
\nonumber
\TVG(\coord{v}_m) - \TVG(\coord{v}_n)
&=& \left|1-\frac{\lambda_m}{|\lambda_{\scriptsize \textrm{max}}|}\right| - \left|1-\frac{\lambda_n}{|\lambda_{\scriptsize \textrm{max}}|}\right| \\
\nonumber
&\overset{(a)}=& \left(1-\frac{\lambda_m}{|\lambda_{\scriptsize \textrm{max}}|}\right) - \left(1-\frac{\lambda_n}{|\lambda_{\scriptsize \textrm{max}}|}\right) \\
&=& \frac{\lambda_n-\lambda_m}{|\lambda_{\scriptsize \textrm{max}}|} > 0,
\end{eqnarray*}
which yields~\eqref{eq:frequency_ordering_real}. Here, equality (a) follows from~\eqref{eq:lambda_max}.
\end{IEEEproof}

\begin{figure}
  \begin{center}
    \subfigure[Ordering of a real spectrum]{\label{fig:spectrum_real}\includegraphics[scale=0.9]{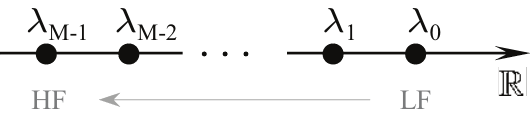}} \\
    \vspace{3mm}
    \subfigure[Ordering of a complex spectrum]{\label{fig:spectrum_complex}\includegraphics[scale=0.9]{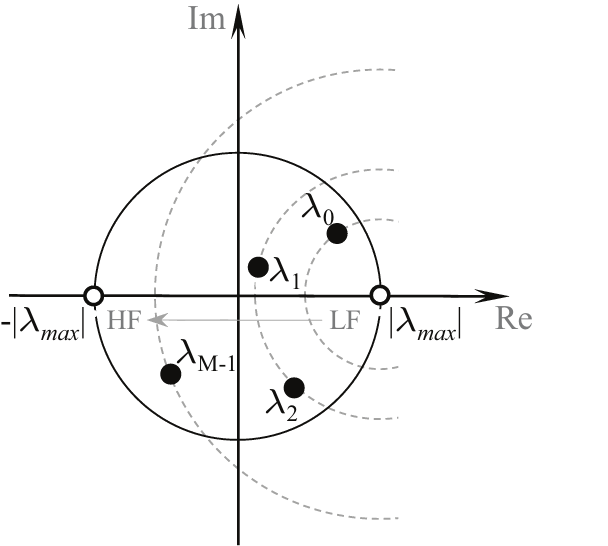}}
  \end{center}
\caption{\label{fig:spectra} Frequency ordering from low frequencies (LF) to high frequencies (HF) for graphs with real and complex spectra.}
\end{figure}

As follows from Theorem~\ref{thm:ordering_real}, if a graph has a real spectrum and its frequencies are ordered as
\begin{equation}
\label{eq:order_real}
\lambda_0 > \lambda_1 > \ldots > \lambda_{M-1},
\end{equation}
then $\lambda_0$ represents the lowest frequency and $\lambda_{M-1}$ is the highest frequency. Moreover, the ordering~\eqref{eq:order_real} is a unique ordering of all frequencies from lowest to highest. This frequency ordering for matrices with real spectra is visualized in Fig.~\ref{fig:spectrum_real}.

The next theorem extends Theorem~\ref{thm:ordering_real} and establishes the relative ordering of two distinct frequencies corresponding to complex eigenvalues.

\begin{theorem}
\label{thm:ordering_complex}
Consider two distinct complex eigenvalues $\lambda_m,\lambda_n\in\C$ of the adjacency matrix $\Adj$. Let $\coord{v}_m$ and $\coord{v}_n$ be the corresponding eigenvectors.
The total variations of these eigenvectors satisfy
\begin{equation}
\label{eq:frequency_ordering_complex}
\TVG(\coord{v}_m) < \TVG(\coord{v}_n)
\end{equation}
if the eigenvalue $\lambda_m$ is located closer to the value $|\lambda_{\scriptsize \textrm{max}}|$ on the complex plane than the eigenvalue $\lambda_n$.
\end{theorem}
\begin{IEEEproof}
This result follows immediately from the interpretation of the total variation~\eqref{eq:TV_eigenvector} as a distance function on the complex plane.
Since $\lambda_{\scriptsize \textrm{max}}\neq 0$ (otherwise $\Adj$ would have been a zero matrix), multiplying both sides of~\eqref{eq:frequency_ordering_complex} by $|\lambda_{\scriptsize \textrm{max}}|$ yields the equivalent inequality
\begin{equation}
\label{eq:frequency_ordering_complex_distance}
\Big| |\lambda_{\scriptsize \textrm{max}}| - \lambda_m \Big| < \Big| |\lambda_{\scriptsize \textrm{max}}| - \lambda_n \Big|.
\end{equation}
The expressions on both sides of~\eqref{eq:frequency_ordering_complex_distance} are the distances from $\lambda_m$ and $\lambda_n$ to $|\lambda_{\scriptsize \textrm{max}}|$ on the complex plane.
\end{IEEEproof}

As follows from Theorem~\ref{thm:ordering_complex}, frequencies of a graph with a complex spectrum are ordered by their distance from $|\lambda_{\scriptsize \textrm{max}}|$. As a result, in contrast to the graphs with real spectra, the induced ordering of complex frequencies from lowest to highest is not unique, since distinct complex frequencies can yield the same total variation for their corresponding frequency components. In particular, all eigenvalues lying on a circle of radius $\rho$ centered at point $|\lambda_{\scriptsize \textrm{max}}|$ on the complex plane have the same total variation $\rho/|\lambda_{\scriptsize \textrm{max}}|$. It also follows from~\eqref{eq:lambda_max} that all graph frequencies $\lambda_m$ can lie only inside and on the boundary of the circle with radius $|\lambda_{\scriptsize \textrm{max}}|$. The frequency ordering for adjacency matrices with complex spectra is visualized in Fig.~\ref{fig:spectrum_complex}.

\begin{figure}
  \begin{center}
    \includegraphics[scale=0.9]{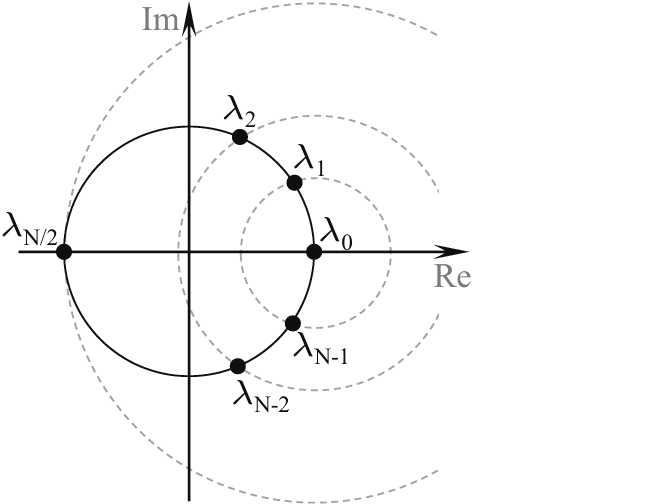}
  \end{center}
\caption{\label{fig:spectrum_time} Frequency ordering for finite discrete periodic time series. Frequencies $\lambda_m$ and $\lambda_{N-m}$ have the same total variations since they lie on the same circle centered around $1$.}
\end{figure}

\mypar{Consistency with DSP theory}
The frequency ordering induced by the total variation~\eqref{eq:TV_graph} is consistent with classical DSP. Recall from~\eqref{eq:exponent} that the cycle graph in Fig.~\ref{fig:graph_time}, which represents finite time series, has a complex spectrum
$$
\lambda_n = e^{-j \frac{2\pi}{N}n}
$$
for $0\leq n < N$. Hence, the total variation of the $n$th frequency component is
\begin{eqnarray*}
\TVG(\coord{v}_n) &=& \left| 1- e^{-j\frac{2\pi n}{N}} \right| \\
&=& \left|1-\cos\frac{2\pi n}{N}\right| + \left|\sin\frac{2\pi n}{N}\right|.
\end{eqnarray*}
Hence, the frequencies $\lambda_m$ and $\lambda_{N-m}$ have the same variation, and the induced order from lowest to highest frequencies is $\lambda_0, \lambda_1, \lambda_{N-1}, \lambda_2, \lambda_{N-2},\ldots$, with the lowest frequency corresponding to $\lambda_0=1$ and the highest frequency corresponding to $\lambda_{N/2}=-1$ for even $N$ or $\lambda_{(N\pm 1)/2}$ for odd $N$. This ordering is visualized in Fig.~\ref{fig:spectrum_time}, and it is the conventional frequency ordering in DSP~\cite{Oppenheim:99}.

\subsection{Frequency Ordering Based on Quadratic Form}
\label{sec:FrequencyQuadratic}
Here we compare our ordering of the frequencies based on the total variation with an ordering based on using the 2-Dirichlet form, $p=2$, like in~\cite{Shuman:13}. Taking $p=2$ in~\eqref{eq:Dirichlet}, we get
\begin{eqnarray}
\nonumber
\Sf_2(\coord{s}) &=& \frac{1}{2} \sum_{n=0}^{N-1} \left| \nabla_n(\coord{s}) \right|^2 \\
\label{eq:quadratic_form}
&=& \frac{1}{2} \left|\left| \coord{s} - \Adjn\coord{s} \right|\right|_2^2 \\
\nonumber
&=& \frac{1}{2} \coord{s}^H \left(\Id- \Adjn\right)^H \left( \Id- \Adjn\right) \coord{s}.
\end{eqnarray}
This quadratic form defines the seminorm
\begin{equation}
\label{eq:seminorm}
\left| \left| \coord{s}  \right| \right|_G = \sqrt{\Sf_2(\coord{s})},
\end{equation}
since $\left(\Id- \Adjn\right)^H \left( \Id- \Adjn\right)$  is a positive-semidefinite matrix. The rationale in~\cite{Shuman:13} is that the quadratic form is small when signal values are close to the corresponding linear combinations of their neighbors' values, and large otherwise.

We introduce an ordering of the graph Fourier basis from lowest to highest frequencies based on the graph shift quadratic form.
As we demonstrate next, this ordering coincides with the ordering induced by the total variation.

The quadratic form~\eqref{eq:quadratic_form} of an eigenvector $\coord{v}$ that corresponds to the eigenvalue $\lambda$ is
\begin{eqnarray}
\nonumber
\Sf_2(\coord{v}) &=& \frac{1}{2} \left|\left| \coord{v} - \Adjn\coord{v} \right|\right|_2^2 \\
\label{eq:form_eigenvector}
&=& \left|1-\frac{\lambda}{|\lambda_{\scriptsize \textrm{max}}|}\right|^2  \left| \left|\coord{v} \right|\right|_2.
\end{eqnarray}

Consider two real eigenvalues $\lambda_m$ and $\lambda_n$ with corresponding eigenvectors $\coord{v}_m$ and $\coord{v}_n$.
If these eigenvalues satisfy $\lambda_m <\lambda_n$, then it follows from~\eqref{eq:form_eigenvector} that
\begin{eqnarray*}
\nonumber
\Sf_2(\coord{v}_m) - \Sf_2(\coord{v}_n)
&=& \left|1-\frac{\lambda_m}{|\lambda_{\scriptsize \textrm{max}}|}\right|^2 - \left|1-\frac{\lambda_n}{|\lambda_{\scriptsize \textrm{max}}|}\right|^2 \\
\nonumber
&=& \left(1-\frac{\lambda_m}{|\lambda_{\scriptsize \textrm{max}}|}\right)^2 - \left(1-\frac{\lambda_n}{|\lambda_{\scriptsize \textrm{max}}|}\right)^2 \\
&=& \frac{\lambda_n-\lambda_m}{|\lambda_{\scriptsize \textrm{max}}|^2}\left( \lambda_m + \lambda_n - 2|\lambda_{\scriptsize \textrm{max}}| \right) > 0.
\end{eqnarray*}
Hence, $\Sf_2(\coord{v}_m) > \Sf_2(\coord{v}_n)$, and we obtain a reformulation of Theorem~\ref{thm:ordering_real} for the graph shift quadratic form.
As a consequence, arranging frequency components in the increasing order of their graph shift quadratic form leads to the same ordering of frequencies~\eqref{eq:frequency_ordering_real} from lowest to highest as the total variation.

A similar reformulation of Theorem~\ref{thm:ordering_complex} for the graph shift quadratic form is demonstrated analogously, which leads to the same ordering of complex frequencies as the ordering induced by the total variation.

\section{Filter Design}
\label{sec:FilterDesign}

When a graph signal is processed by a graph filter, its frequency content changes according to the frequency response~\eqref{eq:frequency_response} of the filter.
Similarly to classical DSP, we can characterize graph filters as low-, high-, and band-pass filters based on their frequency response.

\subsection{Low-, High-, and Band-pass Graph Filters}
\label{sec:LPHP}

Following the DSP convention, we call filters \emph{low-pass} if they do not significantly affect the frequency content of low-frequency signals but attenuate, i.e., reduce, the magnitude of high-frequency signals. Analogously, \emph{high-pass} filters pass high-frequency signals while attenuating low-frequency ones, and \emph{band-pass} filters pass signals with frequency content within a specified frequency band while attenuating all others.

The action of a graph filter $h(\Adj)$ of the form~\eqref{eq:h_A} on the frequency content of a graph signal~$\coord{s}$ is completely specified by its frequency response~\eqref{eq:frequency_response}. For simplicity of presentation, we discuss here graphs with diagonalizable adjacency matrices, for which~\eqref{eq:frequency_response} is a diagonal matrix with $h(\lambda_m)$ on the main diagonal, $0\leq m < M$. In this case, the Fourier transform coefficients of the filtered signal
$$
\coord{\widetilde{s}} = h(\Adj)\coord{s}
$$
are the Fourier transform coefficients of the input signal multiplied element-wise by the frequency response of the filter:
$$
\FT \coord{\widetilde{s}} = \begin{bmatrix}
h(\lambda_0)\\
& \ddots \\
&& h(\lambda_{M-1})
\end{bmatrix}
\coord{\widehat{s}}
=
\begin{bmatrix}
h(\lambda_0)\widehat{s}_0\\
\vdots \\
h(\lambda_{M-1})\widehat{s}_{N-1}
\end{bmatrix}.
$$
Hence, to attenuate the frequency content of a signal inside a specific part of the spectrum, we should design a filter $h(\Adj)$ that for corresponding frequencies $\lambda_m$ satisfies $h(\lambda_m)\approx 0$.

Consider an example of ideal low-pass and high-pass filters $h(\Adj)$ and $g(\Adj)$. Let the cut-off frequency $\lambda_{\scriptsize \textrm{cut}}$ equal to the median of the bandwidth, i.e., be such that exactly half of frequencies $\lambda_m$ are lower frequencies than $\lambda_{\scriptsize \textrm{cut}}$. The frequency responses of these filters are defined as
\begin{equation}
\label{eq:ideal_LP}
h(\lambda_m) = 1-g(\lambda_m) = \begin{cases}
1,&\,\,\, \lambda_m > \lambda_{\scriptsize \textrm{cut}}, \\
0,&\,\,\, \lambda_m \leq \lambda_{\scriptsize \textrm{cut}}.
\end{cases}
\end{equation}

As we demonstrate next, the design of such filters, as well as any low-, high-, and band-pass graph filters, is a linear problem.

\subsection{Frequency Response Design for Graph Filters}
\label{sec:FilterConstruction}

A graph filter can be defined through its frequency response $h(\lambda_m)$ at its distinct frequencies $\lambda_m$, $m=0,\ldots,M-1$.
Since a graph filter~\eqref{eq:h_A} is a polynomial of degree $L$, the construction of a filter with frequency response $h(\lambda_m) = \alpha_m$ corresponds to inverse polynomial interpolation,
i.e., solving a system of $M$ linear equations with $L+1$ unknowns $h_0,\ldots,h_L$:
\begin{eqnarray}
\nonumber
h_0 + h_1\lambda_0 + \ldots + h_{L}\lambda_0^{L} &=& \alpha_0, \\
\nonumber
h_0 + h_1\lambda_1 + \ldots + h_{L}\lambda_1^{L} &=& \alpha_1, \\
\label{eq:system_filter}
&\vdots \\
\nonumber
h_0 + h_1\lambda_{M-1} + \ldots + h_{L}\lambda_{M-1}^{L} &=& \alpha_{M-1}.
\end{eqnarray}
This system can be written as
\begin{equation}
\label{eqn:filterdesign1}
\left[
\!
\begin{array}{cccc}
1&\lambda_0 & \ldots &\lambda_0^{L} \\
1&\lambda_1 & \ldots &\lambda_1^{L} \\
&\vdots&\vdots &\\
1&\lambda_{M-1} & \ldots &\lambda_{M-1}^{L}
\end{array}
\!
\right]
\!\!\!
\left[
\!
\begin{array}{c}
h_0\\
h_1\\
\vdots\\
h_{L}
\end{array}
\!
\right]
\!\!=\!\!
\left[
\!
\begin{array}{c}
\alpha_0\\
\alpha_1\\
\vdots\\
\alpha_{M-1}
\end{array}
\!
\right].
\end{equation}
The system matrix in~\eqref{eqn:filterdesign1} is a full-rank $M\times (L+1)$ Vandermonde matrix~\cite{Lancaster:85,Gantmacher:59}.
Hence, the system has infinitely many exact solutions if $M \leq L$ and one unique exact solution if $M = L+1$.

When $M \geq L+1$, the system is overdetermined and does not have an exact solution. This is a frequent case in practice,
since the number of coefficients in the graph filter may be restricted by computational efficiency or numerical stability requirements.
In this case, we can find an approximate solution, for example, in the least-squares sense.

\begin{figure}
  \begin{center}
    \includegraphics[scale=0.3]{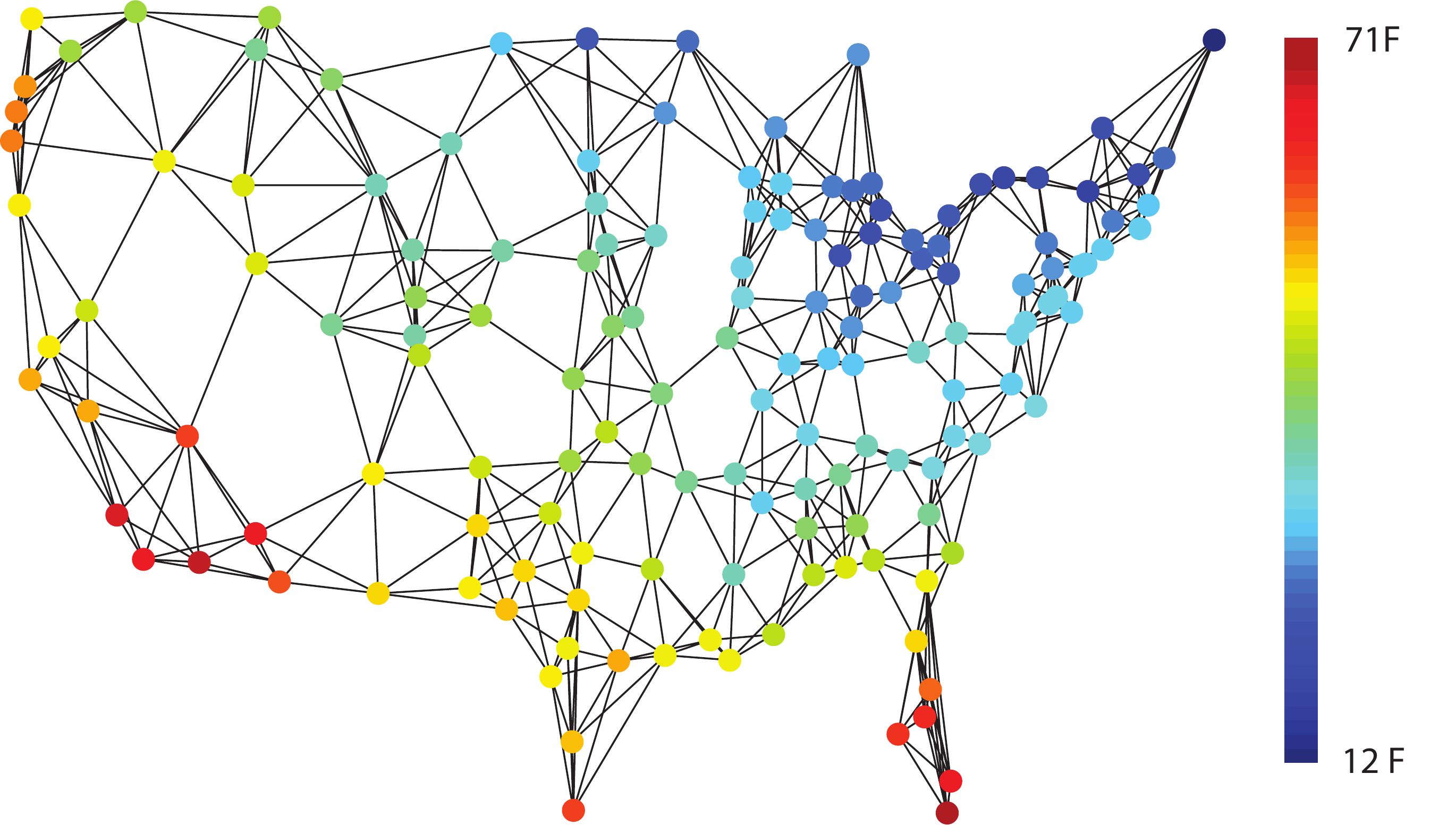}
  \end{center}
\caption{\label{fig:temperature_02012003} Temperature measured by $150$ weather stations across the United States on February 1, 2003}
\end{figure}

As an example of filter construction, consider a network of $150$ weather stations that measure daily temperature near major cities across the United States~\cite{NCDC}.
We represent these stations with a directed $6$-nearest neighbor graph, in which every sensor corresponds to a vertex and is connected to six closest sensors by directed edges.
The edge weight between connected vertices $v_n$ and $v_m$ is
\begin{equation}
\label{eq:A_temperature}
\Adj_{n,m} = \frac{e^{-d_{nm}^2}}{\sqrt{\sum_{k\in\Neighb_n}e^{-d_{nk}^2} \sum_{\ell\in\Neighb_m}e^{-d_{m\ell}^2}}},
\end{equation}
where $d_{n,m}$ denotes the geodesical distance between the $n$th and $m$th sensors. A daily snapshot of all $150$ measurements forms a signal indexed by this graph,
such as the example signal shown in Fig.~\ref{fig:temperature_02012003}.

\begin{figure}
  \begin{center}
    \includegraphics[scale=0.25]{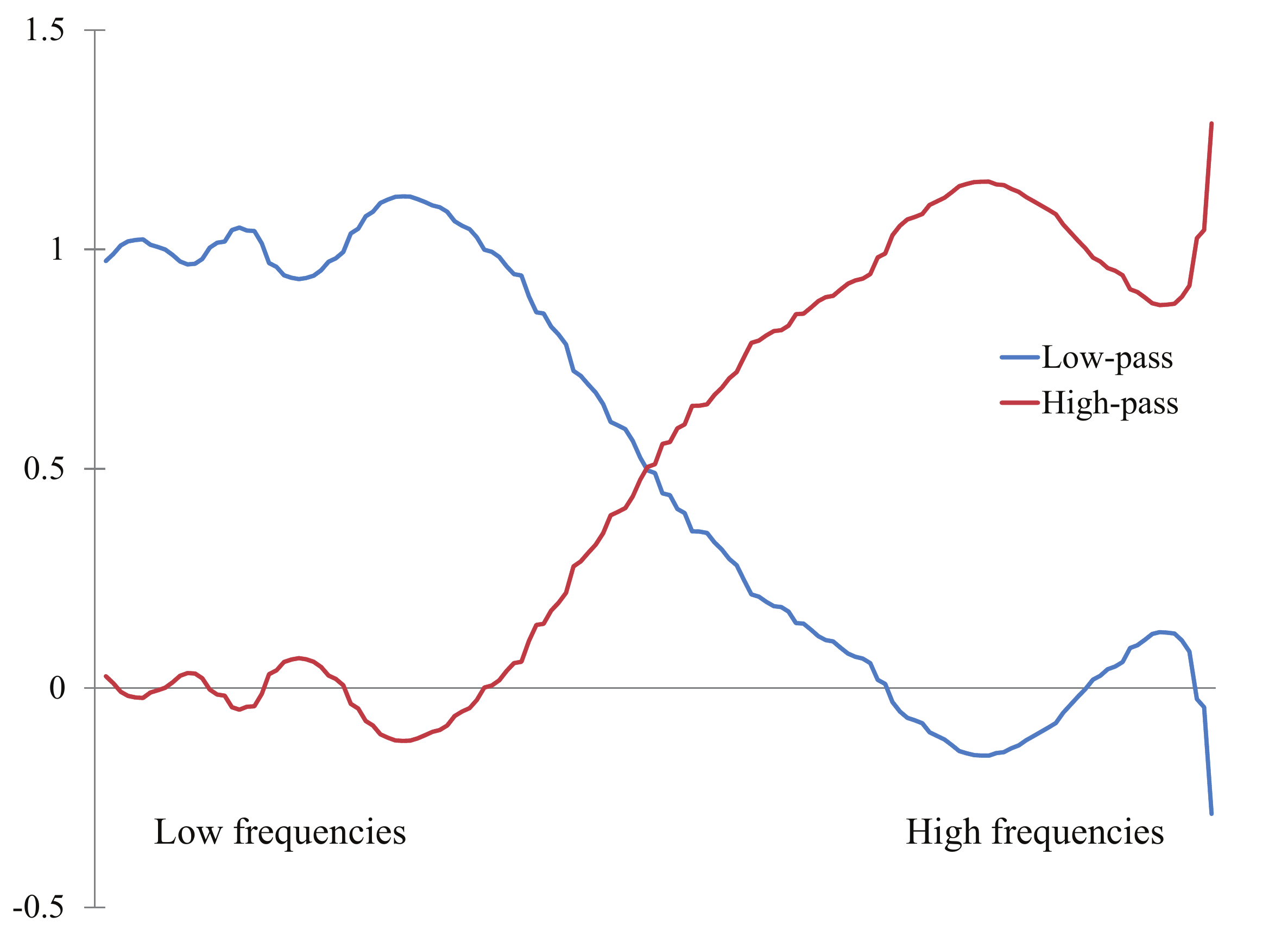}
  \end{center}
\caption{\label{fig:LP_HP} Frequency responses of low-pass and high-pass filters for the sensor graph in Fig.~\ref{fig:temperature_02012003}. The length of the filters is restricted to 10 coefficients.}
\end{figure}

Fig.~\ref{fig:LP_HP} shows the frequency responses of the low- and high-pass filters for this graph that have degree $L=10$.
These filters are least-squares approximations of the ideal low- and high-pass filters~\eqref{eq:ideal_LP}. The frequency response in~\eqref{eq:system_filter} for the low-pass filter is $\alpha_m=1$ for frequencies lower than $\lambda_{\scriptsize \textrm{cut}}$ and $0$ otherwise; and vice versa for the high-pass filter.
By design, the constructed filters satisfy the relation~\cite{Rivlin:69}
\begin{equation}
\label{eq:LP_HP_relation}
h(\Adj) = \Id_N - g(\Adj).
\end{equation}
If we require that $h(\Adj)$ and $g(\Adj)$ do not have the same number of coefficients or if we use an approximation metric other than least squares, the constructed polynomials will not satisfy~\eqref{eq:LP_HP_relation}.

\section{Applications}
\label{sec:Applications}

In this section, we apply the theory discussed in this paper to graphs and datasets that arise in different contexts.
We demonstrate how the \DSPG\ framework extends standard signal processing techniques of band-pass filtering and signal regularization to solve interesting problems in sensor networking and data classification.

\subsection{Malfunction Detection in Sensor Networks}
\label{sec:Sensor}

Today, sensors are ubiquitous. They are usually cheap to manufacture and deploy, so networks of sensors are used to measure and monitor a wide range of physical quantities from structural integrity of buildings to air pollution. However, the sheer quantity of sensors and the area of their deployment may make it challenging to check that every sensor is operating correctly. As an alternative, it is desirable to detect a malfunctioning sensor solely from the data it generates. We illustrate here how the \DSPG\ framework can be used to devise a simple solution to this problem.

\begin{figure}
  \begin{center}
    \includegraphics[scale=0.2]{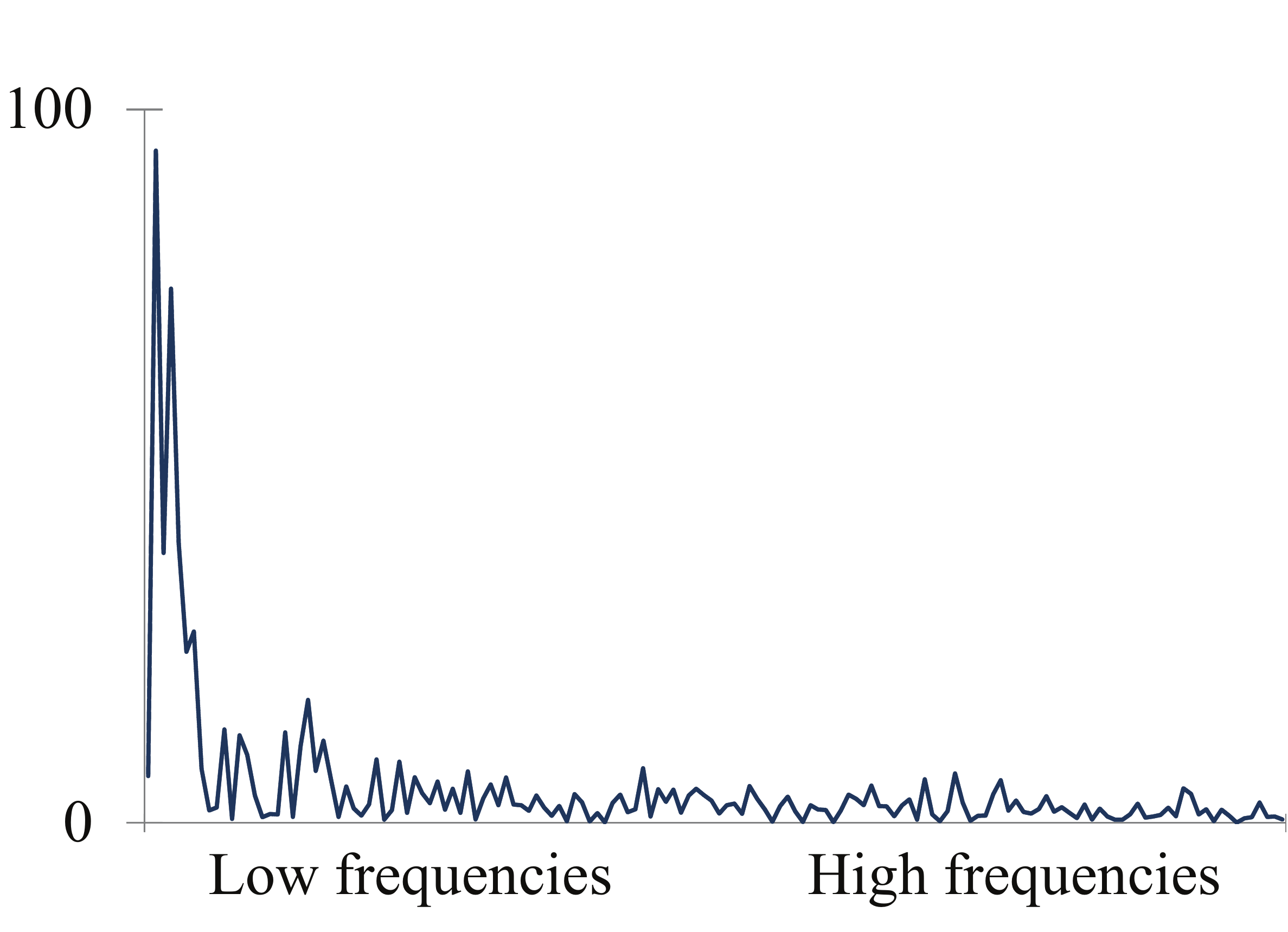}
  \end{center}
\caption{\label{fig:temperature_frequency} The frequency content of the graph signal in Fig.~\ref{fig:temperature_02012003}. Frequencies are ordered from the lowest to highest.}
\end{figure}

Many physical quantities represent graph signals with small variation with respect to the graph of sensors. As an illustration, consider the temperature across the United States measured by $150$ weather stations located near major cities~\cite{NCDC}. An example temperature measurement is shown in Fig.~\ref{fig:temperature_02012003}, and the construction of the corresponding weather station graph is discussed in Section~\ref{sec:FilterConstruction}. The graph Fourier transform of this temperature snapshot is shown in Fig.~\ref{fig:temperature_frequency}, with frequencies ordered from lowest to highest. Most of the signal's energy is concentrated in the low frequencies that have small variation. This suggests that the signal varies slowly across the graph,
i.e., that cities located close to each other have similar temperatures.

\begin{figure}
  \begin{center}
    \subfigure[True measurement] {\label{fig:region_CO_true}\includegraphics[scale=0.28]{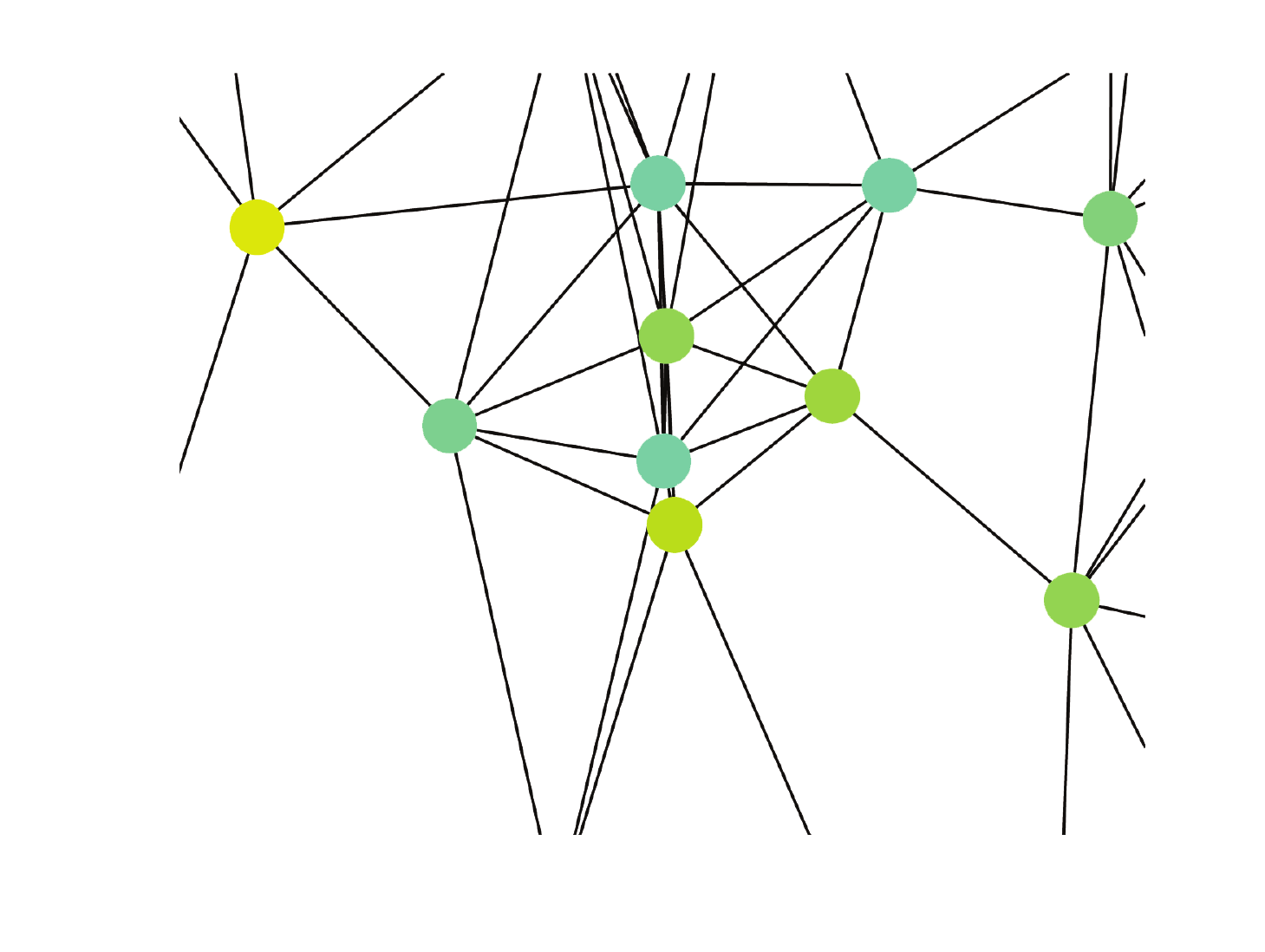}}
    \subfigure[Corrupted measurement]{\label{fig:region_CO_corrupted}\includegraphics[scale=0.28]{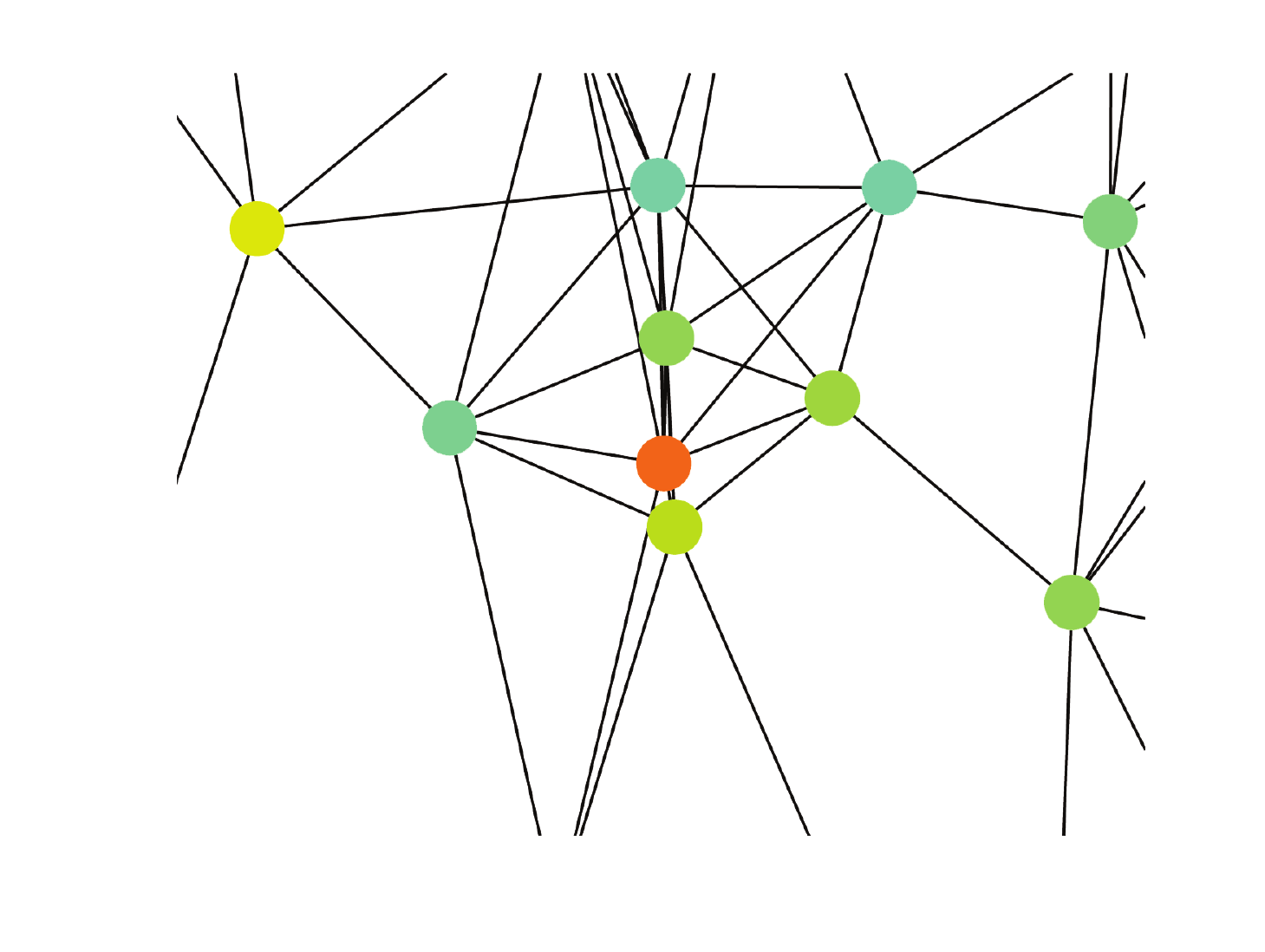}}
  \end{center}
  \vspace{-2mm}
\caption{\label{fig:region_CO} A subgraph of the sensor graph in Fig.~\ref{fig:temperature_02012003} showing the (a) true and (b) corrupted measurement by the sensor located in Colorado Springs, CO.}
\vspace{-5mm}
\end{figure}

A sensor malfunction may cause an unusual difference between its measurements and the measurements of nearby stations. Fig.~\ref{fig:region_CO} shows an example of an (artificially) corrupted measurement,
where the station located near Colorado Springs, CO, reports a temperature that contains an error of~$20$ degrees (temperature at each sensor is color-coded using the same color scheme as in Fig.~\ref{fig:temperature_02012003}). The true measurement in Fig.~\ref{fig:region_CO_true} is very similar to measurements at neighboring cities, while the corrupted measurement in Fig.~\ref{fig:region_CO_corrupted} differs significantly from its neighbors.

Such difference in temperature at closely located cities results in the increased presence of higher frequencies in the corrupted signal.
By high-pass filtering the signal and then thresholding the filtered output, we can detect this anomaly.

\mypar{Experiment}
We consider the problem of detecting a corrupted measurement from a single temperature station. We simulate a signal corruption by changing the measurement of one sensor by~$20$ degrees; such an error is reasonably small and is hard to detect by direct inspection of measurements of each station separately. To detect the malfunction, we extract the high-frequency component of the resulting graph signal using the high-pass filter in Fig.~\ref{fig:LP_HP} and then threshold it. If one or more Fourier transform coefficients exceed the threshold value, we conclude that a sensor is malfunctioning. The cut-off threshold is selected automatically as the maximum absolute value of graph Fourier transform coefficients of the high-pass filtered measurements from the previous three days.

\mypar{Results}
We considered $365$ measurements collected during the year 2003 by all $150$ stations, and conducted $150\times 365=54,750$ tests. The resulting average detection accuracy was $89\%$, so the proposed approach, despite its relative simplicity, correctly detected a corrupted measurement almost $9$ times out of $10$.

\begin{figure}
  \begin{center}
    \subfigure[True signal]{\label{fig:original_HP}\includegraphics[scale=0.18]{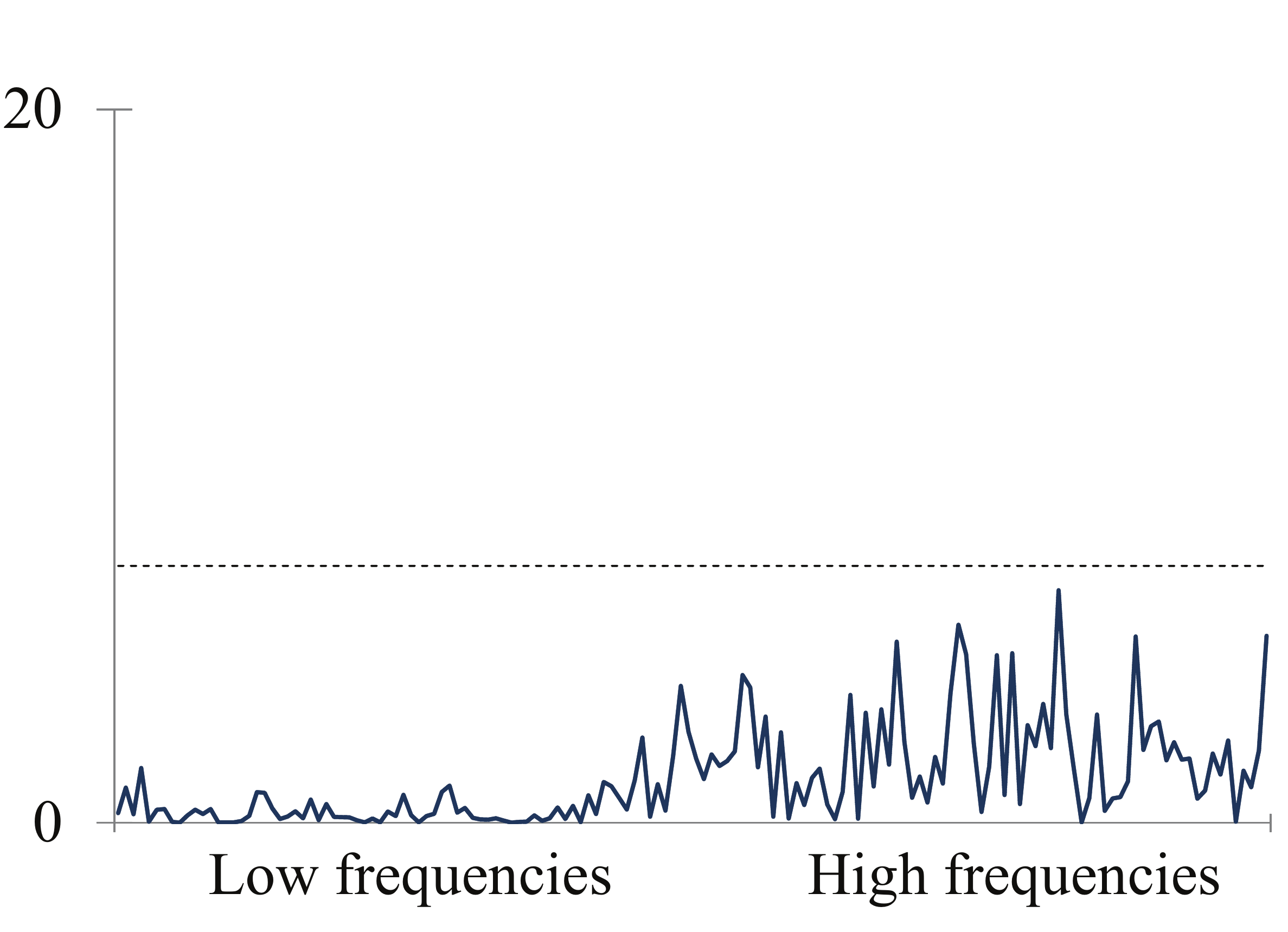}}
    \subfigure[Colorado Springs, CO]{\label{fig:colorado}\includegraphics[scale=0.18]{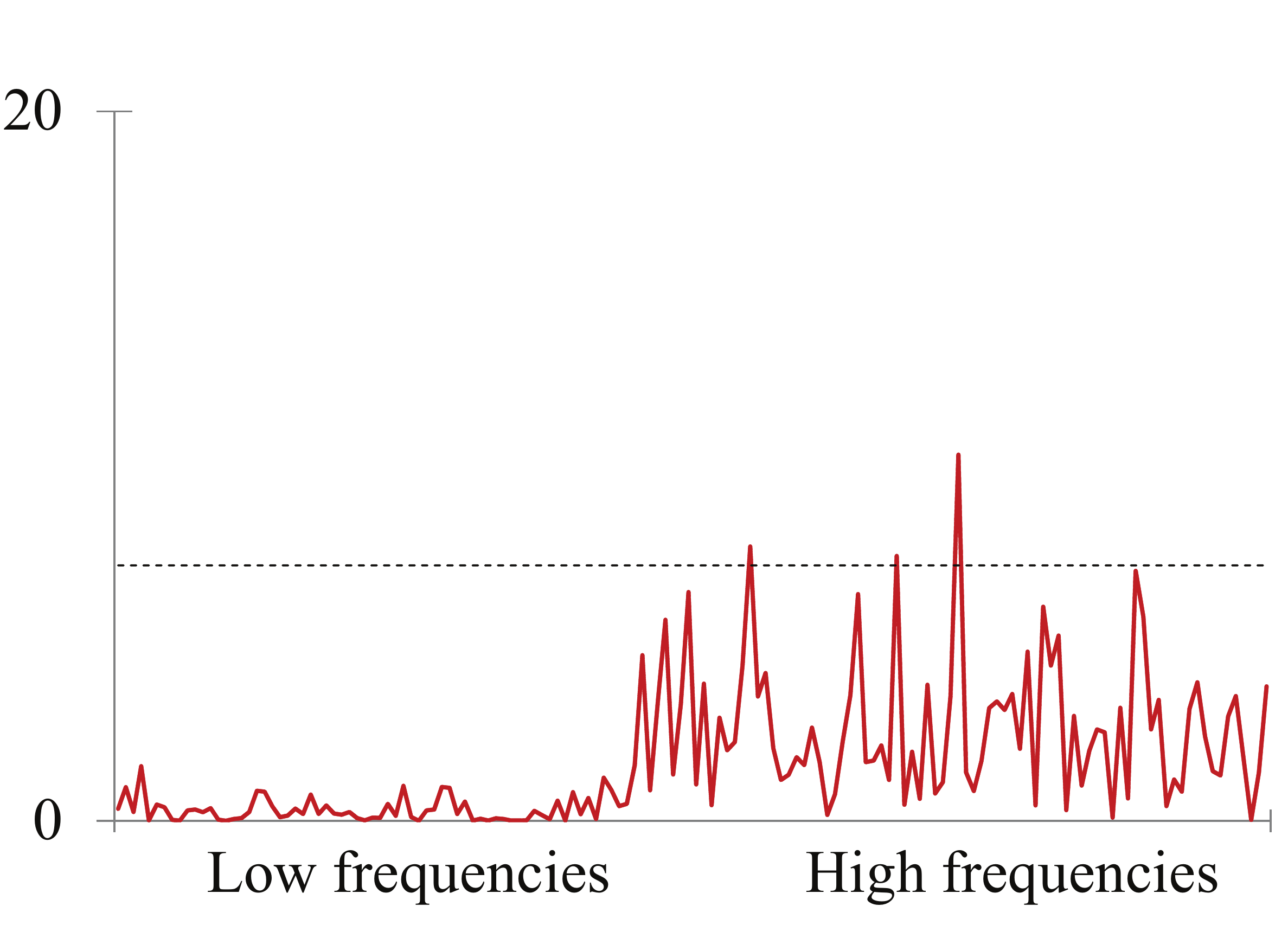}} \\
    \subfigure[Tampa, FL]{\label{fig:florida}\includegraphics[scale=0.18]{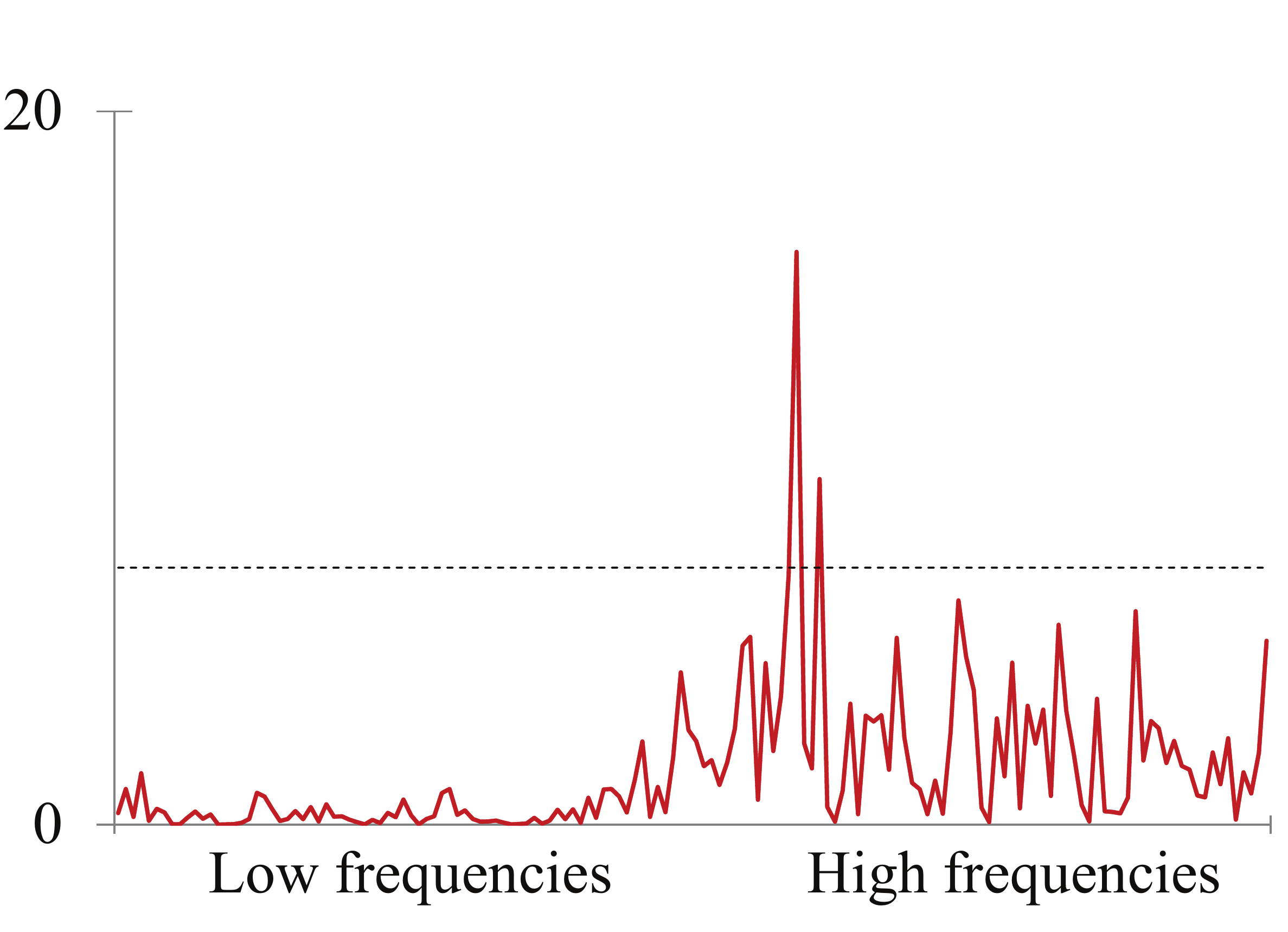}}
    \subfigure[Atlantic City, NJ]{\label{fig:nj}\includegraphics[scale=0.18]{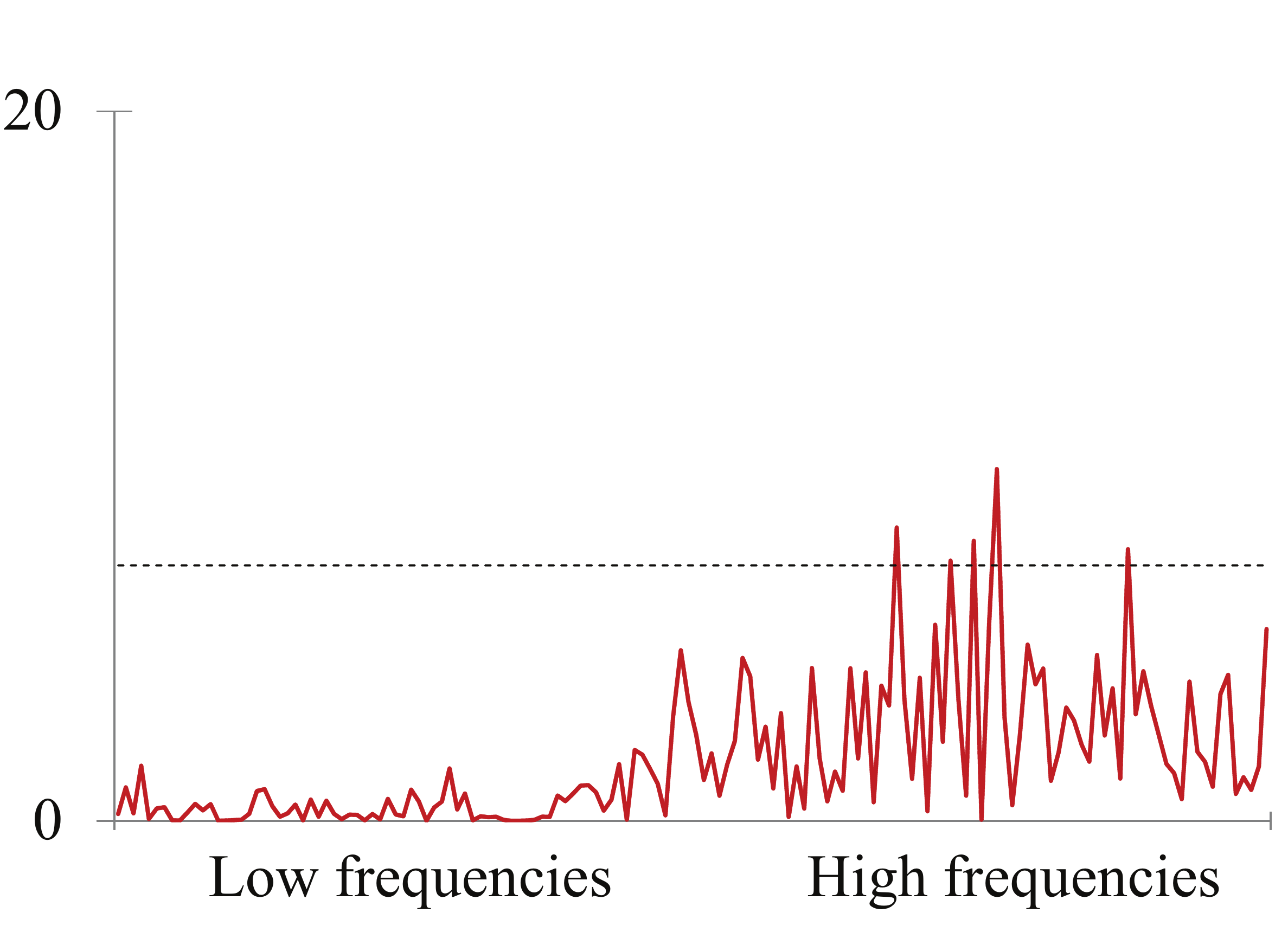}}\\
    \subfigure[Reno, NV]{\label{fig:nevada}\includegraphics[scale=0.18]{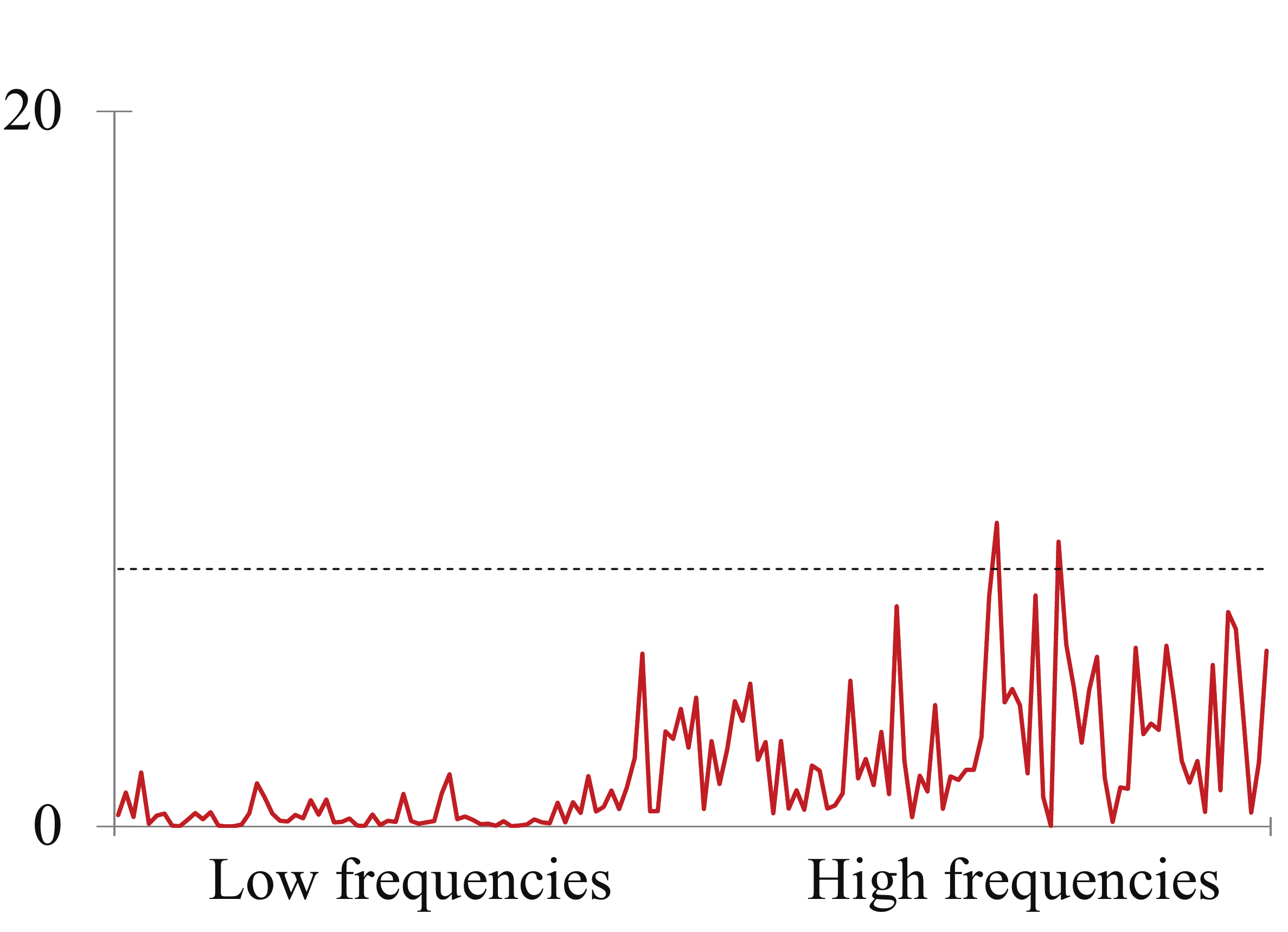}}
    \subfigure[Portland, OR]{\label{fig:oregon}\includegraphics[scale=0.18]{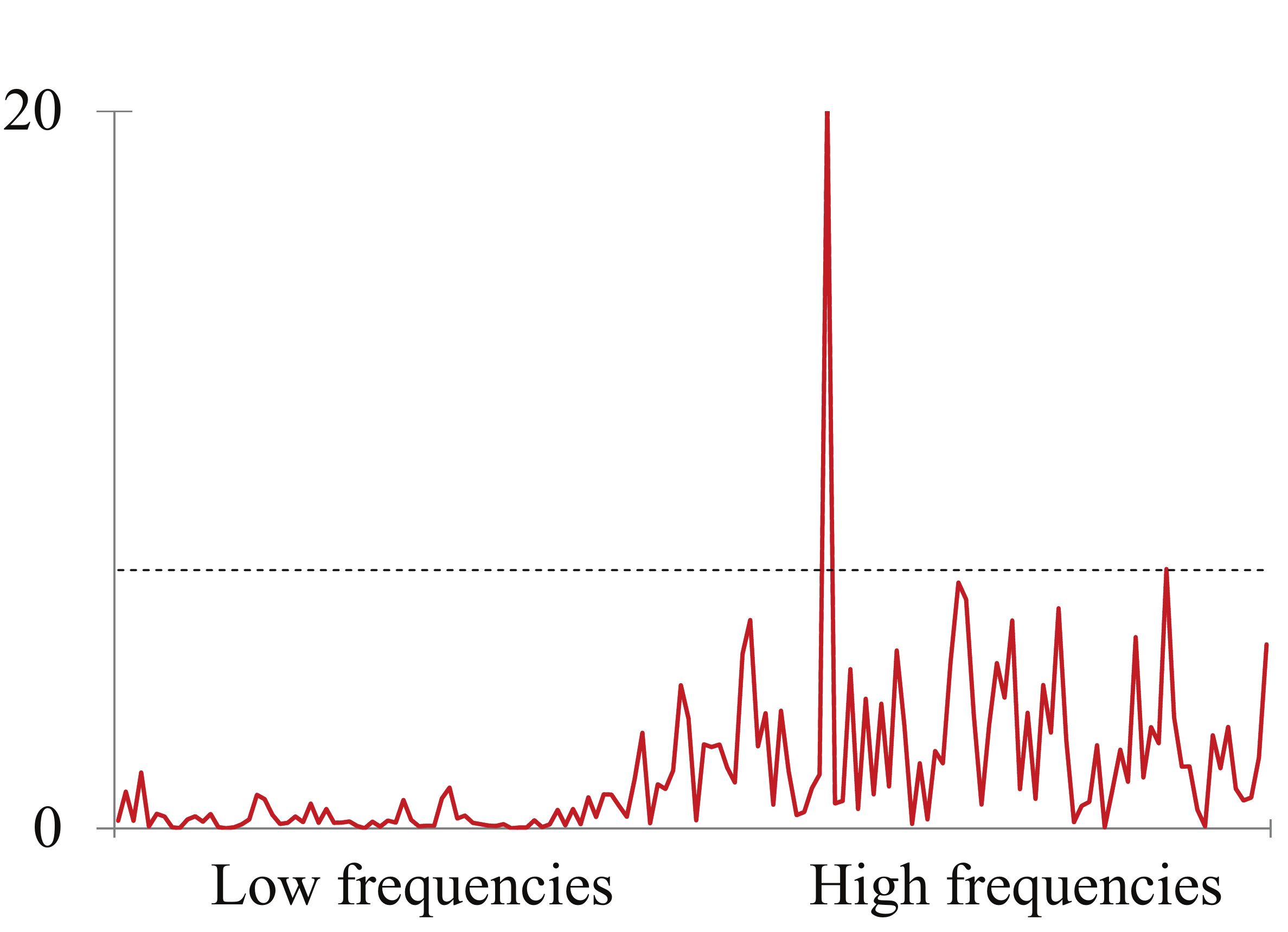}}
  \end{center}
\caption{\label{fig:sensor_malfunction}
The magnitudes of spectral coefficients of the original and corrupted temperature measurements after high-pass filtering: (a) the true signal from Fig.~\ref{fig:temperature_02012003}; (b)-(f) signals obtained from the true signal by corrupting the measurement of a single station located at the indicated city.}
\end{figure}

Fig.~\ref{fig:sensor_malfunction} illustrates the conducted experiment. It shows frequency contents of high-pass filtered signals that contain a corrupted measurement from a sensor at five different locations. A comparison with the high-pass component of the uncorrupted signal in Fig.~\ref{fig:original_HP} shows coefficients above thresholds that lead to the detection of a corrupted signal.

\subsection{Data Classification}
\label{sec:Classification}

Data classification is an important problem in machine learning and data mining~\cite{Duda:00,Chapelle:06}. Its objective is to classify each element of the dataset based on specified characteristics of the data. For example, image and video databases may be classified based on their contents; documents are grouped with respect to their topics; and customers may be distinguished based on their shopping preferences.  In all cases, each dataset element is assigned a label from a pre-defined group of labels.

Large datasets often cannot be classified manually. In this case, a common approach is to classify only a subset of elements and then use a known structure of the dataset to predict the labels for the remaining elements. A key assumption in this approach is that similar elements tend to be in the same class. In this case, the labels tend to form a signal with small variation over the graph of similarities. Hence, information about similarity between dataset elements provides means for inferring unknown labels from known ones.

Consider a graph $G=(\Nodes,\Adj)$ with $N$ vertices that represent $N$ data elements. We assume that two elements are similar to each other if the corresponding vertices are connected; if their connection is directed, the similarity is assumed only in the direction of the edge. We define a signal $\coord{s}^{\scriptsize\textrm{(known)}}$ on this graph that captures known labels. For a two-class problem, this signal is defined as
$$
s^{\scriptsize\textrm{(known)}}_n = \begin{cases}
+ 1,&\,\,\,n\text{th element belongs to class 1,} \\
- 1,&\,\,\,n\text{th element belongs to class 2,} \\
\phantom{-}0,&\,\,\,\text{class is unknown.}
\end{cases}
$$

The predicted labels for all data elements are found as the signal that varies the least on the graph $G=(\Nodes,\Adj)$.
That is, we find the predicted labels as the solution to the optimization problem
\begin{equation}
\label{eq:minimization1}
\coord{s}^{\scriptsize \textrm{(predicted)}} = \underset{\coord{s}\in\R^N}{\argmin}\, \Sf_2(\coord{s})
\end{equation}
subject to
\begin{equation}
\label{eq:condition1}
||\Cm \coord{s}^{{\scriptsize \textrm{(known)}}} - \Cm\coord{s}||_2^2 < \epsilon,
\end{equation}
where $\Cm$ is a $N\times N$ diagonal matrix such that
$$
\Cm_{n,n} = \begin{cases}
1,&\,\,\,\text{if   } s^{{\scriptsize \textrm{(known)}}}_n\neq 0, \\
0,&\,\,\,\text{otherwise.}
\end{cases}
$$
The parameter $\epsilon$ in~\eqref{eq:condition1} controls how well the known labels are preserved.
Alternatively, the problem~\eqref{eq:minimization1} with condition~\eqref{eq:condition1} can be formulated and solved as
\begin{equation}
\label{eq:minimization2}
\coord{s}^{\scriptsize \textrm{(predicted)}} = \underset{\coord{s}\in\R^N}{\argmin}\, \left( \Sf_2(\coord{s}) + \alpha ||\Cm \coord{s}^{{\scriptsize \textrm{(known)}}} - \Cm\coord{s}||_2^2 \right).
\end{equation}
Here, the parameter $\alpha$ controls the relative importance of conditions~\eqref{eq:minimization1} and~\eqref{eq:condition1}.
Once the predicted signal $\coord{s}^{\scriptsize \textrm{(predicted)}}$ is calculated, the unlabeled data elements are assigned to class 1 if $s^{\scriptsize \textrm{(predicted)}}_n > 0$ and another class otherwise.

In classical DSP, minimization-based approaches to signal recovery and reconstruction are called \emph{signal regularization}. They have been used for signal denoising, deblurring and recovery~\cite{Rudin:92, Chan:01, Oliveira:09, Selesnick:13}. In signal processing on graphs, minimization problems similar to~\eqref{eq:minimization1} and~\eqref{eq:minimization2} formulated with the Laplacian quadratic form (see~\eqref{eq:Quadratic_Laplacian} in Appendix~B) are used for data classification~\cite{Zhou:04,Belkin:04,Chapelle:06,Wang:06}, characterization of graph signal smoothness~\cite{Agaskar:13} and recovery~\cite{Shuman:13}. The problems~\eqref{eq:minimization1} and~\eqref{eq:minimization2} minimize the graph shift quadratic form~\eqref{eq:quadratic_form} and represent an alternative approach to graph signal regularization.

\begin{figure}
  \begin{center}
    \includegraphics[scale=0.35]{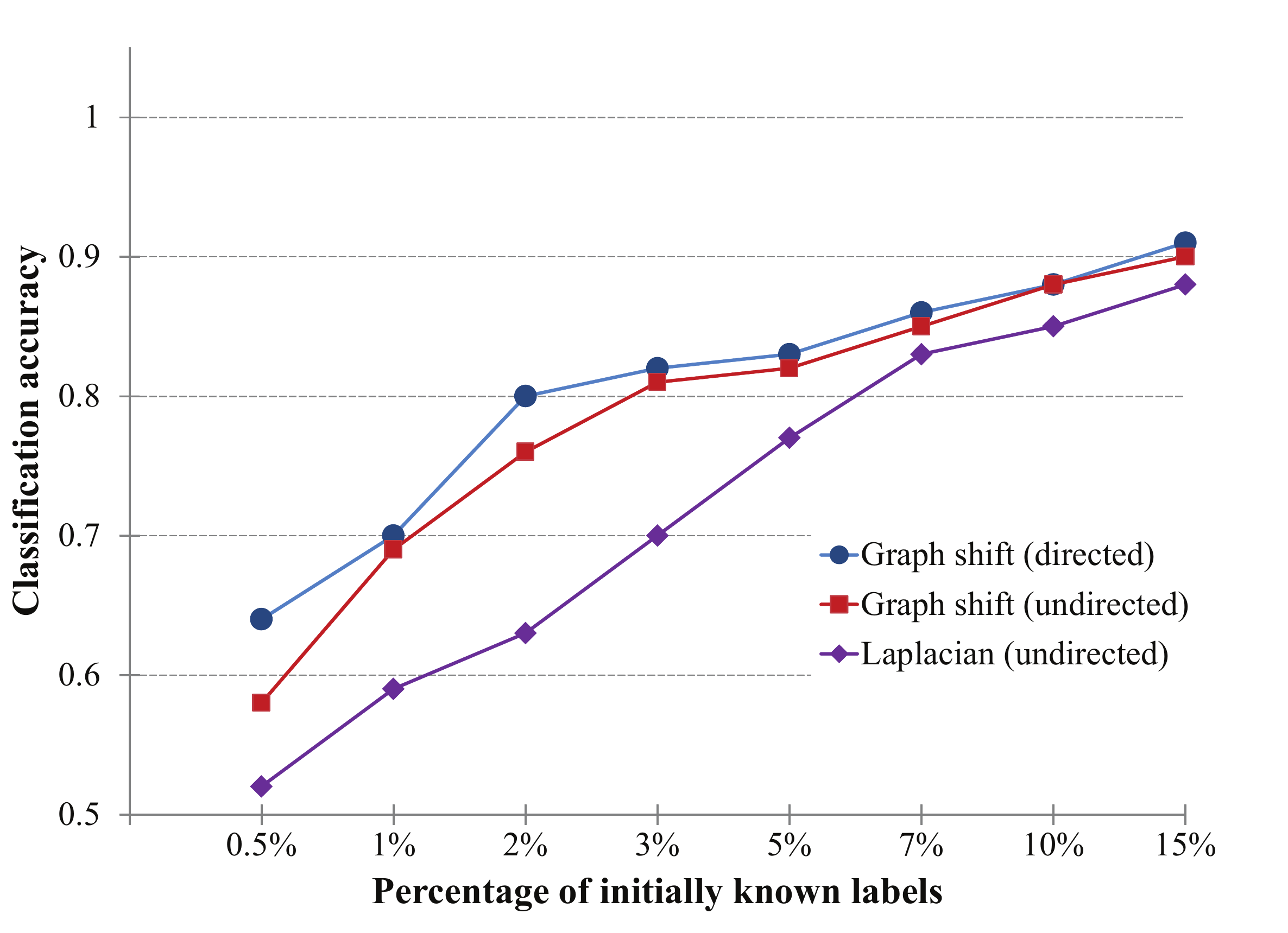}
  \end{center}
  \vspace{-2mm}
\caption{\label{fig:digits_accuracy} Classification accuracy of images of handwritten digits~$0$ and~$1$ using the graph shift-based regularization and Laplacian-based regularization
on weighted and unweighted similarity graphs.}
\end{figure}

\begin{figure}
  \begin{center}
    \includegraphics[scale=0.35]{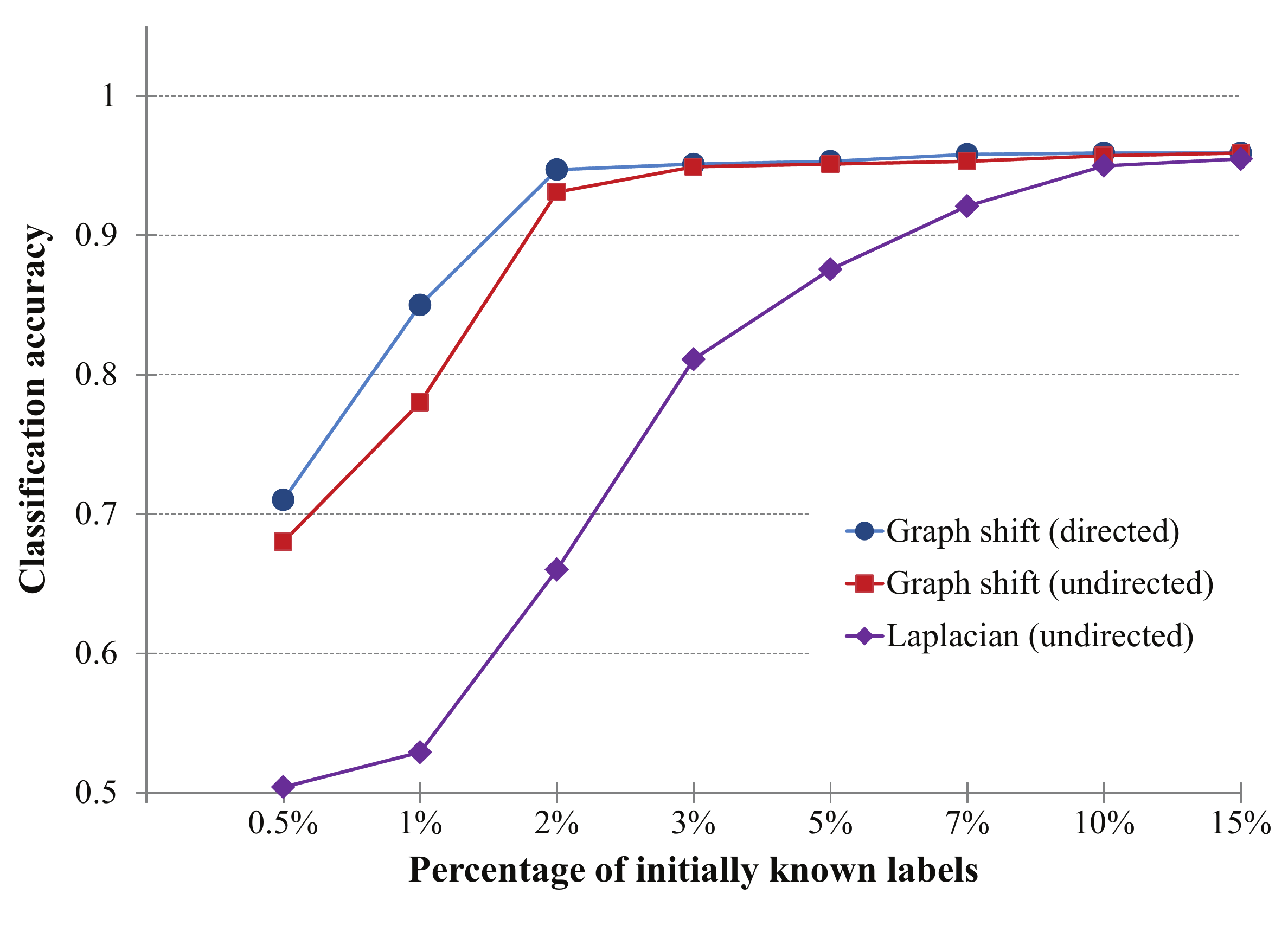}
  \end{center}
  \vspace{-2mm}
\caption{\label{fig:blogs_accuracy} Classification accuracy of political blogs using the graph shift-based regularization and Laplacian-based regularization on an unweighted graph of hyperlink references.}
\end{figure}

\mypar{Experiments}
We illustrate the application of graph signal regularization to data classification by solving the minimization problem~\eqref{eq:minimization2} for two datasets.
The first dataset is a collection of images of handwritten digits~$4$ and $9$~\cite{LeCun:98}. Since these digits look quite similar, their automatic recognition is a challenging task. For each digit, we use $1000$ grayscale images of size $28\times 28$. The graph is constructed by viewing each image as a point in a $28^2 = 784$-dimensional vector space, computing Euclidean distances between all images, and connecting each image with six nearest neighbors by \emph{directed} edges, so the resulting graph is a directed six-nearest neighbor graph.
We consider unweighted graphs,\footnote{
We have also considered weighted graphs with edge weights set to $exp(-d_{n,m}^2)$, where $d_{n,m}$ is the Euclidean distance between the images. This is a common way of assigning edge weights for graphs that reflect similarity between objects~\cite{Belkin:03,Coifman:06,Hammond:11}.
Results obtained for these weighted graphs were practically indistinguishable from the results in
Fig.~\ref{fig:digits_accuracy} and Fig.~\ref{fig:blogs_accuracy} obtained for unweighted graphs.}
for which all non-zero edge weights are set to $1$.

The second dataset is a collection of 1224 online political blogs~\cite{Adamic:05}. The blogs can be either ``conservative'' or ``liberal.'' The dataset is represented by a directed graph with vertices corresponding to blogs and directed edges corresponding to hyperlink references between blogs. For this dataset we also use only the unweighted graph (since we cannot assign a similarity value to a hyperlink).

For both datasets, we consider trade-offs between the two parts of the objective function in~\eqref{eq:minimization2} ranging from $1$ to $100$. In particular, for each ratio of known labels $0.5\%$, $1\%$, $2\%$, $3\%$, $5\%$, $7\%$, $10\%$ and $15\%$, we run experiments for $199$ different values of $\alpha\in\{ 1/100, 1/99,\ldots, 1/2, 1, 2,\ldots, 100\}$, a total of $8\times 199 = 1592$ experiments. In each experiment, we calculate the average classification accuracy over 100 runs. After completing all experiments, we select the highest average accuracy for each ratio of known labels.

For comparison, we also consider the Laplacian quadratic form~\eqref{eq:Quadratic_Laplacian}, and solve the minimization problem
\begin{equation}
\label{eq:minimization_Laplacian}
\coord{s}^{\scriptsize \textrm{(predicted)}} = \underset{\coord{s}\in\R^N}{\argmin}\, \left( \Sf_2^{(L)}(\coord{s}) + \alpha ||\Cm (\coord{s}^{{\scriptsize \textrm{(known)}}} - \coord{s})||_2^2 \right),
\end{equation}
where we vary $\alpha$ between $0.01$ and $100$  as well. Since the Laplacian can only be used with undirected graphs, we convert the original directed graphs for digits and blogs to undirected graphs by making all edges undirected.

For a fair comparison with the Laplacian-based minimization~\eqref{eq:minimization_Laplacian}, we also test our approach~\eqref{eq:minimization2} on the same undirected graphs.
These experiments provide an equal testing ground for the two methods. In addition, by comparing results for our approach on directed and undirected graphs,
we determine whether the direction of graph edges provides additional valuable information that can improve the classification accuracy.

\mypar{Results}
Average classification accuracies for image and blog datasets are shown, respectively, in Fig.~\ref{fig:digits_accuracy} and Fig.~\ref{fig:blogs_accuracy}. For both datasets, the total variation minimization approach~\eqref{eq:minimization2} applied to directed graphs has produced highest accuracies.
This observation demonstrates that using the information about the direction of graph edges improves the classification accuracy of regularization-based classification.

Furthermore, our approach~\eqref{eq:minimization2} significantly outperforms the Laplacian-based approach~\eqref{eq:minimization_Laplacian} on undirected graphs. In particular, for small ratios of known labels, the differences in average accuracies can exceed $10\%$ for image recognition and $20\%$ for blog classification.

\begin{figure}
  \begin{center}
    \subfigure[True labels] {\label{fig:blogs_smooth}\includegraphics[scale=0.35]{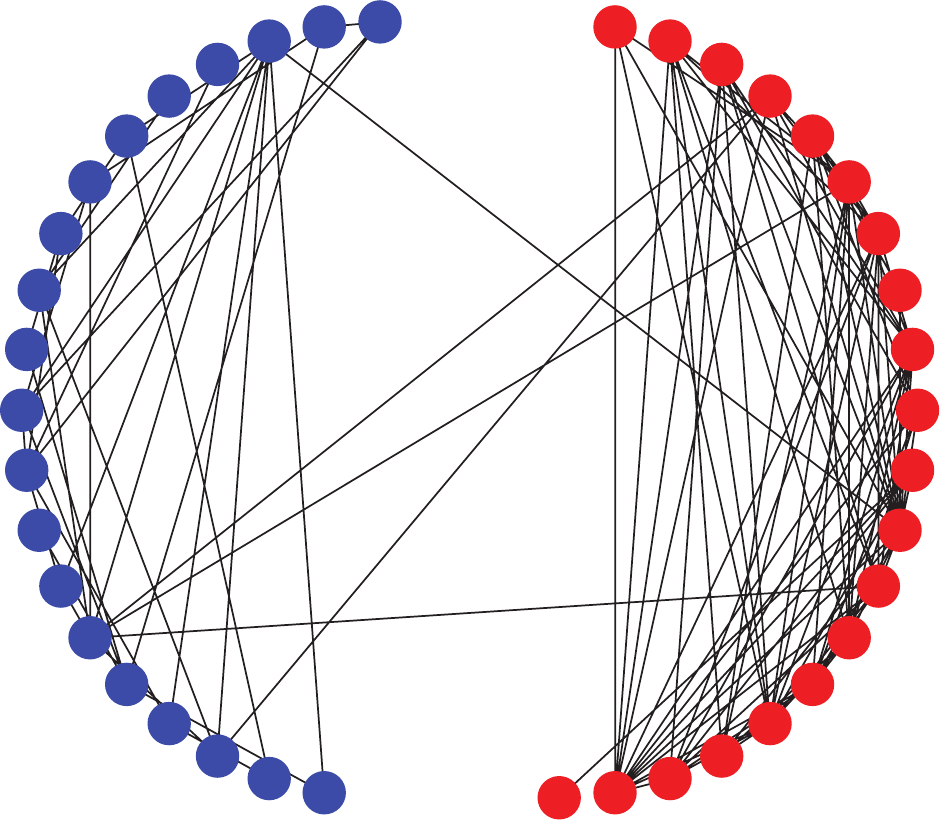}}
    \hspace{5mm}
    \subfigure[Synthetic labels]{\label{fig:blogs_non_smooth}\includegraphics[scale=0.35]{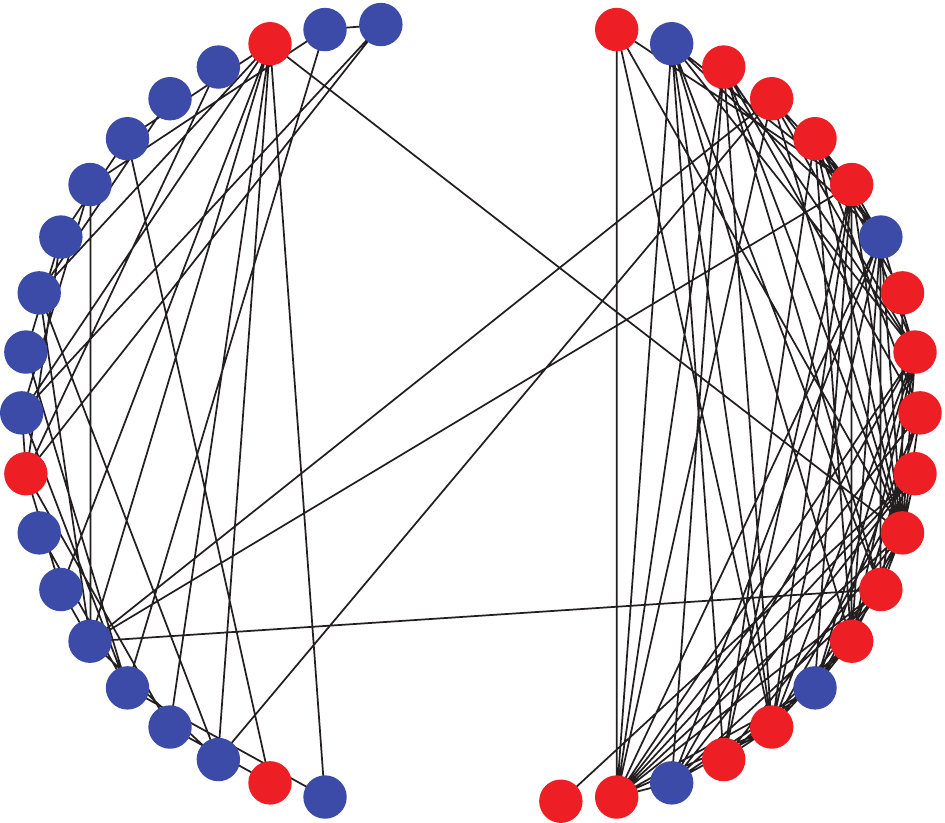}}
  \end{center}
\caption{\label{fig:blogs} A subgraph of~$40$ blogs labels: blue corresponds to ``liberal'' blogs and red corresponds to ``conservative'' ones. Labels in~(a) form a smoother graph signal than labels in~(b).}
\end{figure}

\begin{figure}
  \begin{center}
    \includegraphics[scale=0.3]{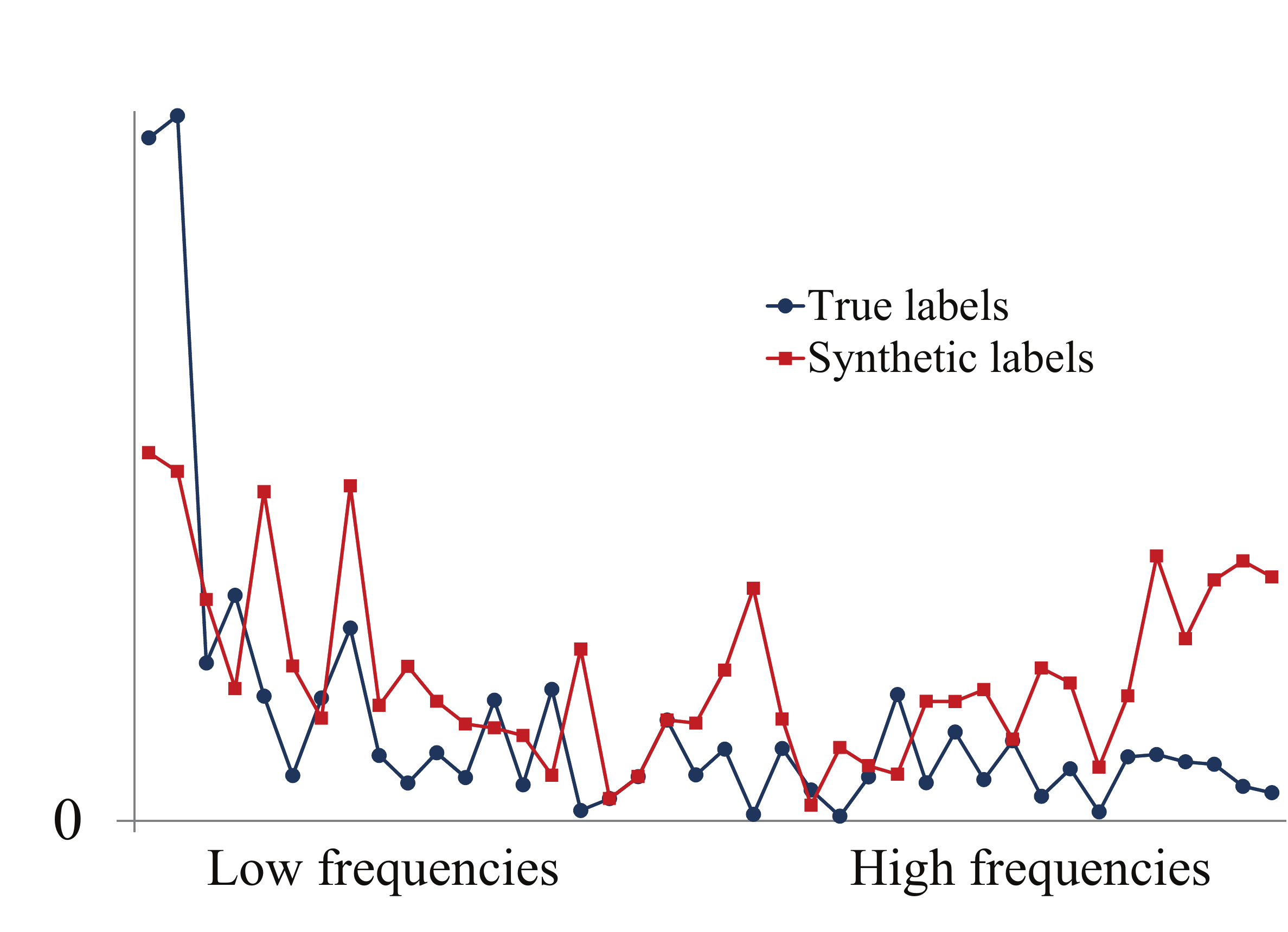}
  \end{center}
\caption{\label{fig:blogs_FT} Magnitudes of the spectral coefficients for graph signals formed by true and synthetic labels in Fig.~\ref{fig:blogs}.}
\end{figure}

\mypar{Discussion}
The following example illustrates how classification based on signal regularization works. Fig.~\ref{fig:blogs} shows a subgraph of~$40$ randomly selected blogs with their mutual hyperlinks. Fig.~\ref{fig:blogs_smooth} contains true labels for these blogs obtained from~\cite{Adamic:05}, while the labels in Fig.~\ref{fig:blogs_non_smooth} are obtained by randomly switching~$7$ out of~$40$ labels to opposite values. The frequency content of the true and synthetic labels as signals on this subgraph are shown in Fig.~\ref{fig:blogs_FT}. The true labels form a signal that has more energy concentrated in lower frequencies, i.e., has a smaller variation than the signal formed by the synthetic labels. This observation supports our assumption that the solution to the regularization problem~\eqref{eq:minimization2} should correspond to correct label assignment.

Incidentally, this assumption also explains why the maximum classification accuracy achieved in our experiments is $96\%$, as seen in Fig.~\ref{fig:blogs_accuracy}. We expect that every blog contains more hyperlinks to blogs of its own type than to blogs of the opposite type. However, after inspecting the entire dataset, we discovered that~$50$ blogs out of $1224$, i.e., $4\%$ of the total dataset, do not obey this assumption. As a result, $4\%$ of all blogs are always misclassified, which results in maximum achievable accuracy of $96\%$.

\section{Conclusions}
\label{sec:Conclusions}
In this paper, we introduced low-, high-, and band-pass signals and low-, high-, and band-pass filters on graphs. These concepts do not have simple and intuitive interpretations for general graphs. We defined them using the concept of frequencies in digital signal processing on graphs. We proposed a novel definition of a total variation on graphs that measures the difference between a graph signal and its shifted version. We then used the total variation to order graph frequencies and to define low- and high-pass graph signals and filters. We demonstrated how to design filters with specified frequency response by finding least squares approximations to solutions of systems of linear algebraic equations. We applied these concepts and methodologies to sensor network analysis and data classification and conducted experiments with real-world datasets of temperature measurements collected by a sensor network and databases of images and hyperlinked documents. Our experimental results showed that the techniques presented in this paper are promising in problems like detecting sensor malfunctions, graph signal regularization, and classification of partially labeled data.

\section*{Appendix~A: Jordan Decomposition}

An arbitrary matrix $\Adj\in\C^{N\times N}$ has $M\leq N$ distinct eigenvalues $\lambda_0,\ldots,\lambda_{M-1}$. Each eigenvalue $\lambda_m$ has $D_m$ corresponding eigenvectors $\coord{v}_{m,0}, \ldots,\coord{v}_{m,D_m-1}$ that satisfy
$$
(\Adj - \lambda_m\Id_N)\coord{v}_{m,d} = 0.
$$
Moreover, each eigenvector $\coord{v}_{m,d}$ can generate a \emph{Jordan chain} of $R_{m,d}\geq 1$ \emph{generalized eigenvectors} $\coord{v}_{m,d,r}$, $0\leq r < R_{m,d}$, where $\coord{v}_{m,d,0}=\coord{v}_{m,d}$, that satisfy
\begin{equation}
\label{eq:generalized_eigenvector_condition}
(\Adj-\lambda_m\Id)\coord{v}_{m,d,r}=\coord{v}_{m,d,r-1}.
\end{equation}
All eigenvectors and corresponding generalized eigenvectors are linearly independent.

For each eigenvector $\coord{v}_{m,d}$ and its Jordan chain of size $R_{m,d}$, we define a \emph{Jordan block} matrix of dimensions~$R_{m,d} \times R_{m,d}$ as
\begin{equation}
\label{eq:Jordan_block}
J_{R_{m,d}}(\lambda_m) = \begin{bmatrix}
\lambda_m & 1 \\
& \lambda_m & \ddots \\
&& \ddots & 1 \\
&&& \lambda_m
\end{bmatrix} \in \C^{R_{m,d}\times R_{m,d}}.
\end{equation}
By construction, each eigenvalue $\lambda_m$ is associated with~$D_m$ Jordan blocks, each of dimension $R_{m,d}\times R_{m,d}$, where $0\leq d < D_m$.
Next, for each eigenvector $\coord{v}_{m,d}$, we collect its Jordan chain into a $N\times R_{m,d}$ matrix
\begin{equation}
\label{eq:generalized_eigenvectors_block}
\Vm_{m,d} = \begin{bmatrix}
\coord{v}_{m,d,0} & \ldots & \coord{v}_{m,d,R_{m,d}-1}
\end{bmatrix}.
\end{equation}
We concatenate all blocks $\Vm_{m,d}$, $0\leq d< D_m$ and $0\leq m<M$, into one block matrix
\begin{equation}
\label{eq:generalized_eigenvectors_matrix}
\Vm = \begin{bmatrix}
\Vm_{0,0} & \ldots & \Vm_{M-1,D_{M-1}}
\end{bmatrix},
\end{equation}
so that the block $\Vm_{m,d}$ is at position $\sum_{k=0}^{m-1}D_k+d$ in this matrix.
Then matrix $\Adj$ is written in its \emph{Jordan decomposition} form as
\begin{equation}
\label{eq:Jordan_decomposition}
\Adj = \Vm \Jm \Vm^{-1},
\end{equation}
where the block-diagonal matrix
\begin{equation}
\label{eq:Jordan_normal_form}
\Jm = \begin{bmatrix}
\Jm_{R_{0,0}}(\lambda_0) \\
& \ddots \\
&& \Jm_{R_{M-1,D_{M-1}}}(\lambda_{M-1})
\end{bmatrix}
\end{equation}
is called the \emph{Jordan normal form} of $\Adj$. The columns of $\Vm$, i.e., all eigenvectors and generalized eigenvectors of $\Adj$, are called the \emph{Jordan basis} of $\Adj$.

\section*{Appendix~B: Connection with Laplacian-based Variation}

The Laplacian matrix for an undirected graph $G=(\Nodes,\Adj)$ with real, non-negative edge weights $\Adj_{n,m}$ is defined as
\begin{equation}
\label{eq:Laplacian}
\Lm = \Dm - \Adj,
\end{equation}
where $\Dm$ is a diagonal matrix with diagonal elements
$$
\Dm_{n,n} = \sum_{m=0}^{N-1}\Adj_{n,m}.
$$
The Laplacian matrix has real non-negative eigenvalues $0=\beta_0 < \beta_1 \leq \beta_2 \leq \ldots \leq \beta_{N-1}$ and a complete set of corresponding orthonormal eigenvectors $\coord{u}_n$ for $0\leq n < N$.

Similarly to the \DSPG\ definition of the graph Fourier transform~\eqref{eq:graph_FT}, the Laplacian-based Fourier transform expands a graph signal $\coord{s}$ into the eigenbasis of $\Lm$~\cite{Shuman:13}. The total variation is defined as
\begin{equation}
\label{eq:TV_Laplacian}
\TVL(\coord{s})  = \sum_{n=0}^{N-1} \left( \sum_{m\in\Neighb_n} \Adj_{n,m}\left(s_n-s_m \right)^2\right)^{1/2},
\end{equation}
and the graph Laplacian quadratic form is
\begin{equation}
\label{eq:Quadratic_Laplacian}
\Sf_2^{(L)}(\coord{s})  = \coord{s}^T \Lm \coord{s}.
\end{equation}
In particular, the Laplacian quadratic form~\eqref{eq:TV_Laplacian} of a Fourier basis vector is
\begin{equation}
\label{eq:TVL_eigenvector}
\Sf_2^{(L)}(\coord{u}_n) = \beta_n.
\end{equation}
It imposes the following order of the Laplacian Fourier basis from the lowest frequency to the highest one:
\begin{equation}
\label{eq:order_Laplacian}
\coord{u}_0,\coord{u}_1,\ldots,\coord{u}_{N-1}.
\end{equation}

For a general graph, the total variation~\eqref{eq:TV_graph} and the graph shift quadratic form~\eqref{eq:quadratic_form}
are different from the~\eqref{eq:TV_Laplacian} and ~\eqref{eq:Quadratic_Laplacian}.
However, as we demonstrate in the following theorem, the \DSPG\ and the Laplacian-based approach to signal processing on graphs lead to the same graph Fourier basis, notions of low and high frequencies,
and frequency ordering on any regular\footnote{All vertices of a $d$-regular graph have the same degree $d$, so that $\sum_{m=0}^{N-1}\Adj_{n,m}=d$.} graph.

\begin{theorem}
\label{thm:Lapl_equivalence}
The quadratic forms~\eqref{eq:quadratic_form} and~\eqref{eq:Quadratic_Laplacian} induce the same ordering on the graph Fourier basis
for regular graphs.
\end{theorem}
\begin{IEEEproof}
Consider a $d$-regular graph with adjacency matrix $\Adj$. Since the Laplacian matrix~\eqref{eq:Laplacian} can be defined only for undirected graphs with real non-negative edge weights,
we also require that $\Adj=\Adj^T$ and has only real non-negative entries. Hence, $\Adj$ has real eigenvalues and a complete set of orthonormal eigenvectors~\cite{Lancaster:85},
and its Jordan decomposition~\eqref{eq:Jordan_decomposition} becomes the eigendecomposition
$$
\Adj = \Vm \Eig \Vm^T,
$$
where $\Eig$ is the diagonal matrix of eigenvalues.
Since the graph is $d$-regular, its Laplacian matrix~\eqref{eq:Laplacian} satisfies
$$
\Lm=d\Id-\Adj = \Vm(d\Id - \Eig)\Vm^T.
$$
Hence, $\Lm$ and $\Adj$ have the same eigenvectors $\coord{u}_n=\coord{v}_n$, i.e., the same graph Fourier basis.
The corresponding eigenvalues are $\beta_m=d-\lambda_m$. Since the smallest eigenvalue of $\Lm$ is $\beta_0=0$, we also obtain $\lambda_{\scriptsize \textrm{max}}=d$.

The graph shift quadratic form~\eqref{eq:quadratic_form} of the eigenvector $\coord{v}_m$ satisfies
\begin{eqnarray}
\nonumber
\Sf_2(\coord{u}_m)
&=&  \frac{1}{2}\left| \left|\left(\Id - \frac{1}{d}\Adj \right) \coord{u}_m  \right|\right|_2^2 \\
\nonumber
&=&  \frac{1}{2}\left(1 - \frac{\lambda_m}{d} \right)^2  \\
\nonumber
&=&  \frac{1}{2d^2}\left(d - \lambda_m \right)^2  \\
\nonumber
&=&  \frac{1}{2d^2}\beta_m^2  \\
\label{eq:form_relation}
&=&  \frac{1}{2d^2} \left( \Sf_2^{(L)}(\coord{s}) \right)^2.
\end{eqnarray}
Since $\beta^2/(2d^2)$ is a monotonically increasing function for $\beta\geq 0$, it follows from~\eqref{eq:form_relation} that ordering the graph Fourier basis $\coord{u}_n$, $0\leq n < N$, by increasing values of the quadratic form~\eqref{eq:quadratic_form} leads to the same order as~\eqref{eq:order_Laplacian}. Hence, the notions of low and high frequencies, and frequency orderings from lowest to highest coincide on regular graphs for the \DSPG\ and the Laplacian-based approach.
\end{IEEEproof}

\bibliographystyle{IEEEbib}
\begin{small}
\bibliography{references}
\end{small}
\end{document}